\documentclass[letterpaper,twocolumn,superscriptaddress,
	    nobibnotes,aps,prb,longbibliography,floatfix]{revtex4-1}
\pdfoutput=1
\usepackage[utf8]{inputenc}
\usepackage[T1]{fontenc}
 
\usepackage{amsmath,amssymb,amsfonts}
\usepackage{amstext, mathrsfs, textcomp}
\usepackage{nicefrac}
\usepackage{multirow}

\usepackage{hyperref}
\usepackage{subfigure}
\usepackage{graphicx}
\usepackage{xcolor}

\graphicspath{{Figures/}{TOC_ACS/}}

\newcommand{\FIGCAPTIONPREFIX}{}

\newcommand{\TITLE}{Evolutionary Multi-Objective Optimisation of Colour Pixels based on Dielectric Nano-Antennas}

\begin{document}

\title{\TITLE}

\author{\firstname{Peter R.} \surname{Wiecha}}
\email[e-mail~: ]{peter.wiecha@cemes.fr}
\affiliation{CEMES-CNRS, Universit\'e de Toulouse, CNRS, UPS, Toulouse, France}

\author{\firstname{Arnaud} \surname{Arbouet}}
\email[e-mail~: ]{arnaud.arbouet@cemes.fr}
\affiliation{CEMES-CNRS, Universit\'e de Toulouse, CNRS, UPS, Toulouse, France}

\author{\firstname{Christian} \surname{Girard}}
\affiliation{CEMES-CNRS, Universit\'e de Toulouse, CNRS, UPS, Toulouse, France}

\author{\firstname{Aur\'elie} \surname{Lecestre}}
\affiliation{LAAS-CNRS, Universit\'e de Toulouse, CNRS, INP, Toulouse, France}

\author{\firstname{Guilhem} \surname{Larrieu}}
\affiliation{LAAS-CNRS, Universit\'e de Toulouse, CNRS, INP, Toulouse, France}

\author{\firstname{Vincent} \surname{Paillard}}
\email[e-mail~: ]{vincent.paillard@cemes.fr}
\affiliation{CEMES-CNRS, Universit\'e de Toulouse, CNRS, UPS, Toulouse, France}

\begin{abstract}
The rational design of photonic nanostructures consists in anticipating their optical response from simple models and their systematic variations.
This strategy, however, has limited success when multiple objectives are simultaneously targeted because it requires demanding computational schemes.
To this end, evolutionary algorithms can drive the morphology of a nano-object towards an optimum through several cycles of selection, mutation and cross-over, mimicking the process of natural selection. 
Here, we present a numerical technique to design photonic nanostructures with optical properties optimised along several arbitrary objectives. 
We combine evolutionary multi-objective algorithms with frequency-domain electro-dynamical simulations to optimise the design of colour pixels based on silicon nanostructures that resonate at two user-defined, polarisation-dependent wavelengths. 
The scattering spectra of optimised pixels fabricated by electron beam lithography show excellent agreement with the targeted objectives. 
The method is self-adaptive to arbitrary constraints, and therefore particularly apt for the design of complex structures within predefined technological limits.
\end{abstract}

\maketitle

Over the last decade, the field of nanophotonics or nano-optics has been rapidly increasing, mainly driven by plasmonics, since noble metal nanoparticles allow to spectrally tune plasmon resonances\cite{tan_plasmonic_2014} and tailor several optical properties like directional scattering,\cite{lindfors_imaging_2016} polarisation conversion,\cite{black_optimal_2014} optical chirality\cite{valev_chirality_2013} or nonlinear effects\cite{kauranen_nonlinear_2012}.
Recently, high-index dielectric nanostructures have gained increasing interest thanks to their ability to provide exceptionally strong electric\cite{albella_electric_2014, bakker_magnetic_2015} and magnetic\cite{ginn_realizing_2012, kuznetsov_magnetic_2012, schmidt_dielectric_2012} resonances, tunable from the UV to the near IR.\cite{cao_tuning_2010, traviss_antenna_2015, zhao_full-color_2016}
In analogy to plasmonics, it is possible to design functionalities like transmissive metasurfaces,\cite{yu_high-transmission_2015} enhanced nonlinear effects\cite{shcherbakov_enhanced_2014, wiecha_enhanced_2015} or directional scattering.\cite{fu_directional_2013}

\begin{figure*}[tb]
\centering
\includegraphics[page=1]{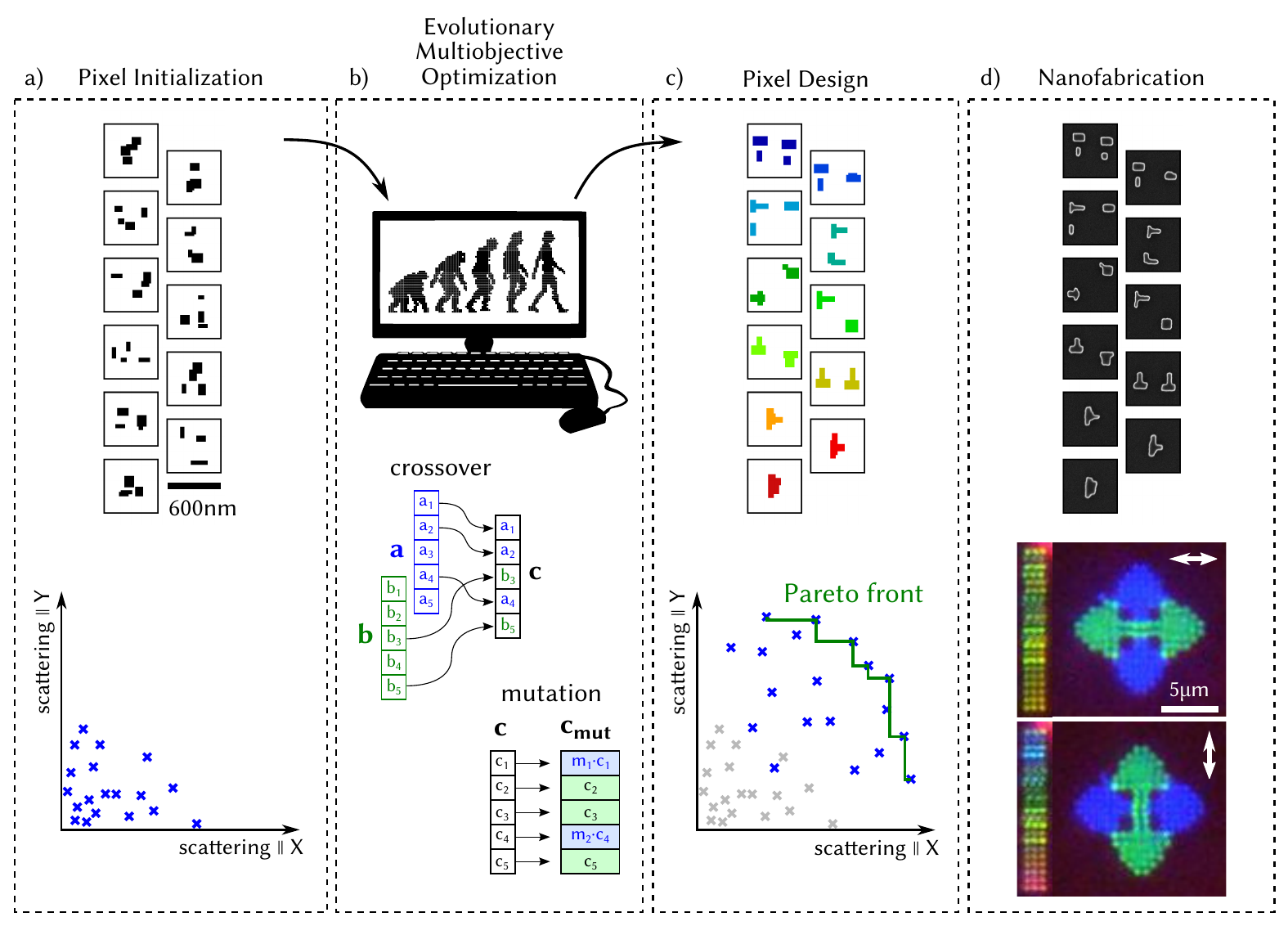} 
\caption{
\textbf{Illustration of evolutionary multi-objective optimisation (EMO).}
a) Randomized initialisation of the pixel-population for the EMO algorithm. The fitness of the individuals is weak and does not yet contain the actual set of non-dominated solutions.
b) Evolution of the pixel-population: In a reproduction step, new solutions to the given problem are generated by mixing of (cross-over step) and allowing random changes to (mutation step) the parameters defining the individuals of the former generation. The population of the new generation is finally obtained after an evaluation and a selection procedure.
c) Optimum pixel-population at the end of the evolution cycle: In case of convergence, the population evolved towards the set of individuals forming the Pareto front. These non-dominated solutions cannot be further optimised in all objectives simultaneously.
d) Nanofabrication and characterization of the polarisation dependent colour-pixels. Top: SEM images of selected structures. Center and bottom: \(X\)-, respectively \(Y\)-polarisation filtered darkfield images of pictograms composed by the EMO-designed nanostructures.
}
\label{fig:sketch_optimization}
\end{figure*}

When designing photonic nanostructures, a particular geometry is usually selected from qualitative considerations and its properties are subsequently studied systematically. 
As it comes to applications, a more convenient approach is to define the requested properties and design a nanostructure that optimally exhibits the desired features.
For the latter approach, a structure model has to be developed, which, based on a certain set of parameters, can describe a large variety of particle geometries.
However, this leads to huge parameter spaces which usually cannot be explored systematically.
Also trial-and-error is not an efficient search strategy.
More promising techniques are evolutionary optimisation strategies which, by mimicking natural selection, are able to find fittest parameter sets to a complex non-analytical problem. \cite{sivanandam_introduction_2008}

In the field of nanophotonics, evolutionary algorithms have been applied to the maximisation of field enhancement,\cite{forestiere_particle-swarm_2010, feichtner_evolutionary_2012, forestiere_genetically_2012, forestiere_inverse_2016} scattering from plasmonic particles\cite{ginzburg_resonances_2011, macias_heuristic_2012} or to the design of hybrid plasmonic/dielectric antennas\cite{bigourdan_design_2014, mirzaei_superscattering_2014}.
Such methods were also successfully used on more technological applications like electron-beam field emission sources,\cite{chen_optimal_2007} waveguide couplers \cite{mengyu_wang_optimization_2014} or core-shell nanoparticles for hyperthermia.\cite{kessentini_particle_2011}

These studies were limited to the maximisation of one target property at a specific wavelength and polarisation.
Such single-objective scenarii represent the simplest case of an optimisation problem, while a structure that concurrently matches multiple objectives will be in general more difficult to design. 
In a recent work, genetic multi-objective optimisation was used on plasmonic waveguides. 
A figure of merit describing the waveguide and its robustness against geometrical variations were maximised simultaneously.\cite{jung_robust_2016}
Evolutionary multi-objective optimisation (EMO) strategies\cite{deb_multi-objective_2001} could lead to considerable improvements in the design of 
wavelength dependent (multi-)directional scattering,\cite{shegai_bimetallic_2011} multiresonant antennas \cite{aouani_ultrasensitive_2013} or polarisation dependent tailored optical behaviour\cite{dopf_coupled_2015}. 
Nanoantennas possessing multiple resonances, for instance at the fundamental and harmonic frequencies, may also be optimised by EMO to enhance nonlinear effects or fluorescence spectroscopy.\cite{giannini_excitation_2008,harutyunyan_enhancing_2012,celebrano_mode_2015}

In this paper, we present a combination of EMO with the Green Dyadic Method (GDM) for self-consistent full-field electro-dynamical simulations\cite{girard_near_2005}.
We apply the EMO-GDM technique to design dielectric (silicon) nanoantennas that concurrently maximise the scattering at different wavelengths, dependent on the polarisation of the incident light.
Finally, from the outcome of the EMO, we fabricated Si nanostructures on a silicon-on-insulator (SOI) substrate and measured their optical response by confocal darkfield scattering microscopy, yielding an excellent agreement with the optimisation predictions.
Possible applications of such nano-scatterers are holographic colour-filters\cite{zhao_full-color_2016} or colour rendering and printing close to the diffraction limit. 
The latter has been demonstrated either using plasmonic\cite{tan_plasmonic_2014, goh_comparative_2016} or dielectric nanostructures\cite{cao_tuning_2010}.
Polarisation dependent, dual-colour pixels have been recently reported using plasmonic nanoapertures.\cite{li_dual_2016} 
While plasmonic nanoantennas provide widely tunable single mode responses from simple geometries (pillars in Ref.~\onlinecite{tan_plasmonic_2014}, cuboids in Ref.~\onlinecite{goh_comparative_2016} and crosses in Ref.~\onlinecite{li_dual_2016}), 
dielectric nanostructures often support high order and degenerate modes in a narrow spectral range.
Therefore, an EMO scheme is of particular interest for the design of multiresonant dielectric nanostructures.

\subsection*{Evolutionary optimisation of scattering efficiency}

\begin{figure}[tb]
\centering
\includegraphics[page=1]{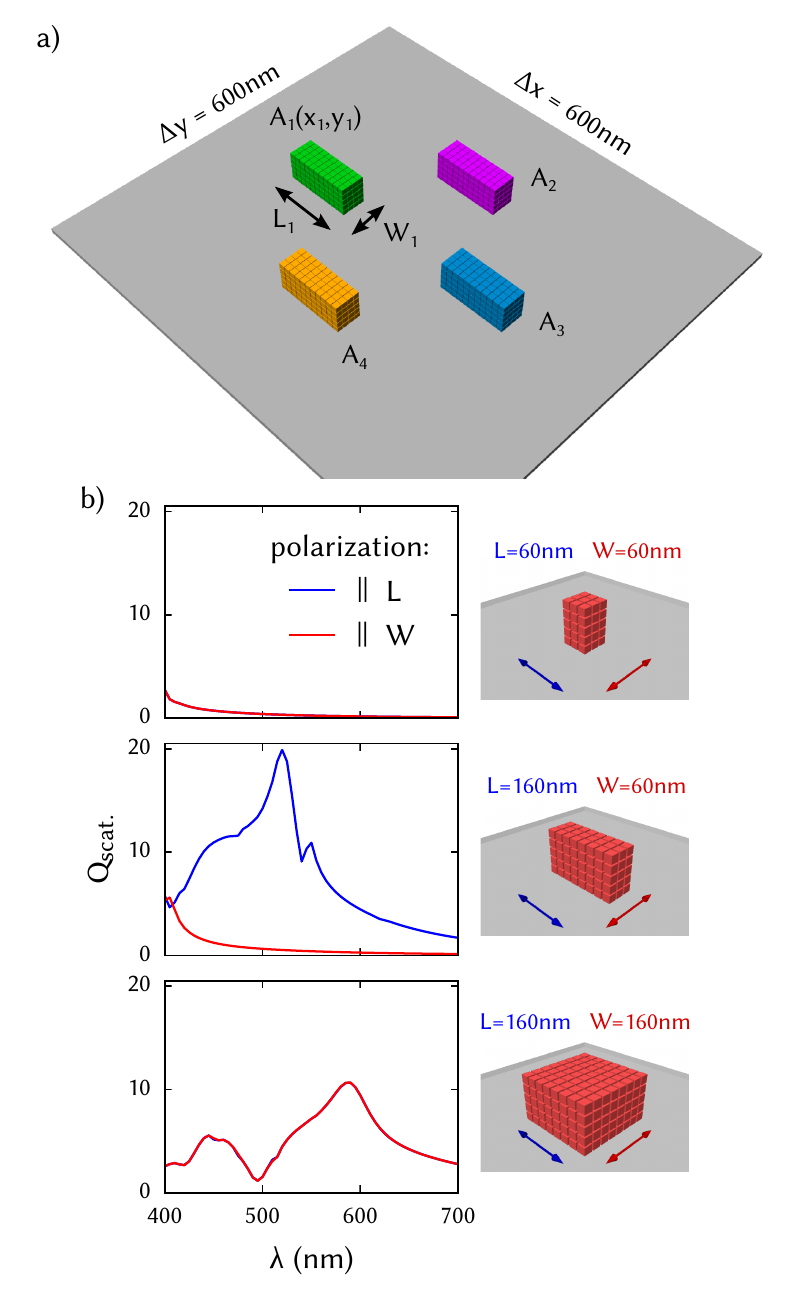} 
\caption{\textbf{Structure model for evolutionary multi-objective optimisation.}
a) Example of silicon block arrangement forming a pixel.
b) Scattering efficiencies calculated for individual silicon blocks of 
the minimum allowed size (top), minimum width and maximum length (center), and maximum possible size (bottom).
The constraints are \(L_{\text{min.}}=W_{\text{min.}}=60\,\)nm and \(L_{\text{max.}}=W_{\text{max.}}=160\,\)nm, the height is fixed to \(H=100\,\)nm.}
\label{fig:sketch_optimization_problem}
\end{figure}

Evolutionary optimisation numerical techniques mimic the selection process which drives the evolution of species in nature. 
Each individual of an initial population is first evaluated along one (single-objective optimisation) or several (multi-objective optimisation) figures of merit summarized in a so-called fitness function. 
Only the best individuals are selected and the next generation is obtained from a reproduction procedure, as illustrated in figure~\ref{fig:sketch_optimization}b. 
The new generation undergoes the same evaluation-selection-reproduction process and, after several cycles, individuals with optimised properties are obtained.

In the present study, we propose to maximise simultaneously the scattering efficiencies \(Q_{\text{scat}}\) of silicon nanoantennas at a first wavelength \(\lambda_X\) for an incident polarisation along \(X\) and at a second wavelength \(\lambda_Y\) for an incident polarisation along \(Y\).
\(Q_{\text{scat}}\) is defined as the ratio between the scattering cross-section \(\sigma_{\text{scat}}\) and the geometrical cross-section of a nanostructure.
We therefore consider a set (or ``population'') of nanostructures which are compared using a fitness function consisting of the scattering efficiencies \(Q_{\text{scat}}(\lambda_X)\) and \(Q_{\text{scat}}(\lambda_Y)\). 
The initial generation consists in a collection of \(N\) antennas with randomly initialised designs. 
At each optimisation step, the scattering efficiencies of all the antennas are compared and the geometries yielding the maximum \(Q_{\text{scat}}(\lambda_X)\) and \(Q_{\text{scat}}(\lambda_Y)\) are selected to generate the set of \(N\) antennas used for the next iteration. 
For more informations about the determination of fitness and selection in multi-objective optimisation see Ref.~\onlinecite{beume_sms-emoa:_2007}. 
This process of evaluation, selection and reproduction is repeated until the maximum number of iterations is reached. 
At the end of the optimisation process a set of optimal solutions called the Pareto front is obtained (see figure~\ref{fig:sketch_optimization}c). 
These optimal (or non-dominated) solutions cannot be further optimised in one of the objectives (increasing \(Q_{\text{scat}}(\lambda_X)\) for instance), without worsening the other target value (decreasing \(Q_{\text{scat}}(\lambda_Y)\) for instance). 
The convergence towards the Pareto front during the optimisation process is illustrated in {\color{blue}the supporting informations (SI), section~D}.

For the electro-dynamical simulations, we use a volume integral technique in the frequency domain, namely the Green Dyadic method\cite{martin_generalized_1995} ({\color{blue}see Methods})

The ``population'' of antenna morphologies to be considered in the computation must be diverse enough to explore, after several generations, a significant fraction of possible solutions.
However, this requires a model with a large number of parameters, significantly slowing down convergence.
Furthermore, the optimised geometries must remain within the limits of fabrication capabilities, hence have neither too many nor too small features.
For these reasons we use a very simple model, based on four individual silicon elements with variable dimensions and positions, placed on a SiO\(_2\) substrate (\(n \approx 1.5\)) within a limited area of \(600\,\times 600\,\)nm\(^2\). 
A sketch of the model is shown in Fig.~\ref{fig:sketch_optimization_problem}a.

Both, the \(x\)- and \(y\)-dimension of each block is allowed to vary between \(60\,\)nm and \(160\,\)nm in steps of \(20\,\)nm, corresponding to the precision of a state of the art electron-beam lithographic system.
The height \(H\) is fixed to \(100\,\)nm, equal to the silicon overlayer thickness of our SOI substrate.
Overlapping antennas are allowed, corresponding antennas are fused together  resulting in a maximum possible size of \(320 \times 320\,\)nm\(^2\).
The constrained area ensures that the planewave excitation in our simulations is a good approximation for the illumination loosely focused through a low-NA dark field objective ({\color{blue}see also Methods}).

Spectra of single silicon-cuboids with dimensions corresponding to the given size-limits are shown in figure~\ref{fig:sketch_optimization_problem}b. 
For simplicity, the positions are discretised in steps of \(20\)\,nm. In order to validate this large stepsize, we calculated spectra for the same structures using different discretisation stepsizes, which yielded comparable results ({\color{blue}see SI, Sec.~B}).

Finally, we note that the number of possible parameter combinations in this model is larger than \( 1 \times 10^{15}\) ({\color{blue}see SI, Sec.~A}).
We conclude that it is inconceivable to use a brute-force strategy (evaluation of all possible combinations). In view of the large parameter space, the convergence and reproducibility of the evolutionary optimisations have been carefully checked by addressing the stability of the predicted geometries with respect to the number of cycles and to different initial populations, respectively ({\color{blue}see SI, section~D}).

\begin{figure*}[t]
\centering
\includegraphics[page=1]{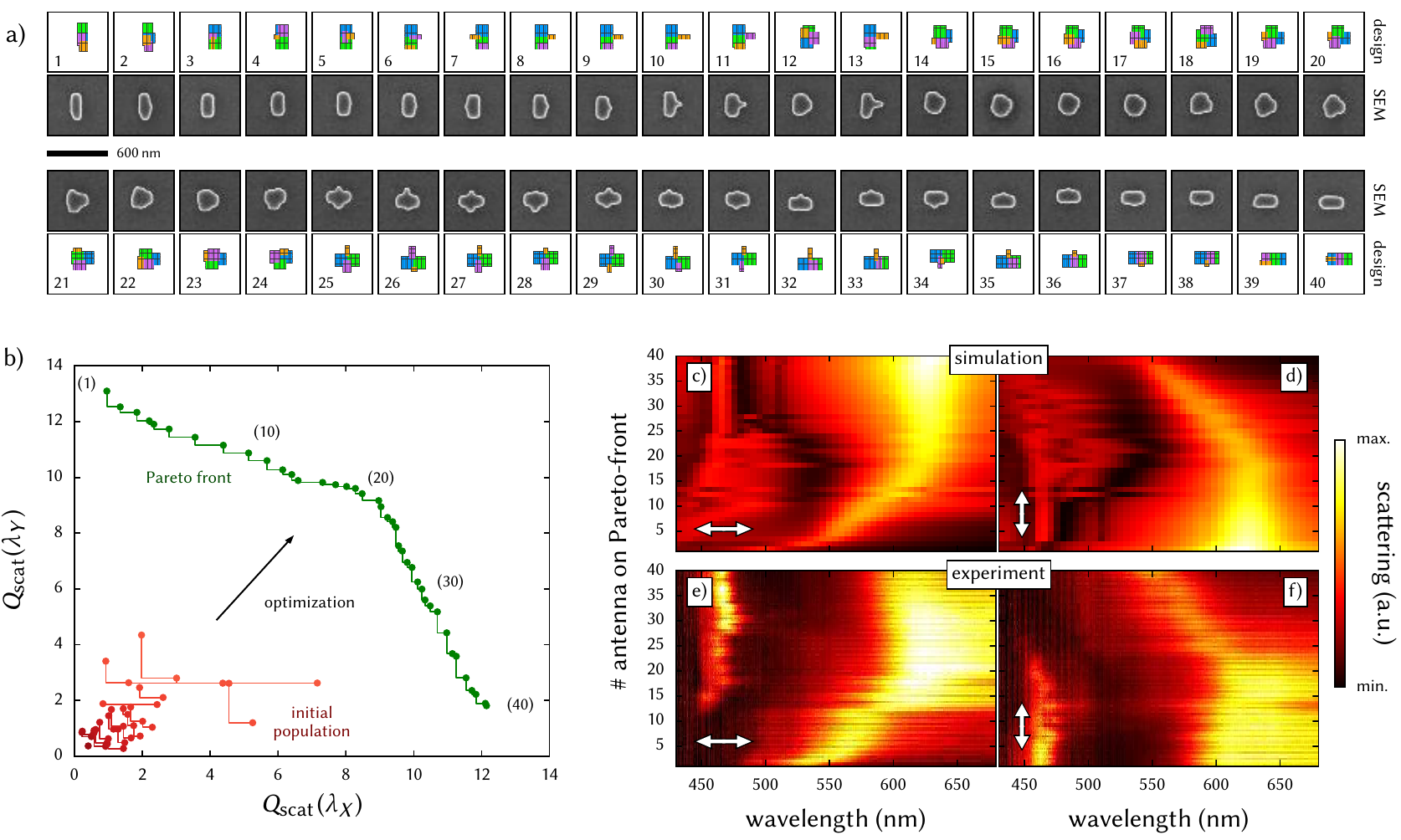}
\caption{\textbf{Results of evolutionary multi-objective optimisation for identical target wavelengths \(\lambda_X = \lambda_Y = 630\,\)nm.}
(a) Structures of the Pareto front and corresponding SEM images. 
All fields are \(600 \times 600\,\text{nm$^{2}$}\) large.
Blue, green, purple and orange dots are used to highlight the positions of the sub-blocks the structures consist of.
(b) Pareto front (green) and randomized initial population (red).
(c-f) Scattering spectra for \(X\) (left) and \(Y\) polarisation (right). Simulated and experimental spectra are shown in the top and bottom rows, respectively.
}
\label{fig:pareto_630_630}
\end{figure*}

\subsection*{Proof of principle: \(\lambda_X=\lambda_Y\)}

In a first step, we test the EMO-GDM technique on a simple problem. 
A single target wavelength \(\lambda_{\text{max.}} = 630\,\)nm is selected, at which \(Q_{\text{scat}}\) is maximised simultaneously for \(X\) and \(Y\) polarisation.
The structures of the final population and the corresponding Pareto front after an evolution over \(200\) generations are shown in Fig.~\ref{fig:pareto_630_630}a and~\ref{fig:pareto_630_630}b. 
The geometries of the initial population are compared to those on the Pareto front in the {\color{blue}SI, Sec.~C}.

The geometries found by evolutionary optimisation are then transformed into a lithographic mask, which we use to produce the silicon nanostructures on a SOI substrate ({\color{blue}see Methods}). 
Fig.~\ref{fig:pareto_630_630}a shows a comparison of the design with SEM images of the sample.
Simulated (Fig.~\ref{fig:pareto_630_630}c-d) and experimental spectra (Fig.~\ref{fig:pareto_630_630}e-f) are in very good agreement. 
We note that the higher order resonance around \(\lambda=450\,\)nm for structures at the edge of the Pareto front is enhanced in the experiment compared to the simulations. 
This is due to a cavity effect in the SiO\(_2\)-layer of the SOI substrate, which is not taken into account in the GDM simulations ({\color{blue}see SI, Sec.~I}).

The outermost individuals on the Pareto front (particles (1) and (40)) correspond to equivalent results of a single-objective optimisation using one target wavelength and polarisation. 
We observe in these cases, that all four sub-antennas are combined during the evolution to form a single rod-like antenna along the target polarisation direction.
In agreement with literature, this yields an optimum scattering efficiency with respect to the considered polarisation direction (``1'': \(Y\), ``40'': \(X\)) -- at the expense of a very low scattering for the perpendicular polarisation.\cite{traviss_antenna_2015}
To obtain comparably high scattering efficiencies for both polarisations (particle ``20'' and neighbours), the evolution produces cross-like antennas.

\subsection*{Evolutionary optimisation of double resonant nanostructures}

\begin{figure}[t]
\centering
\includegraphics[page=1]{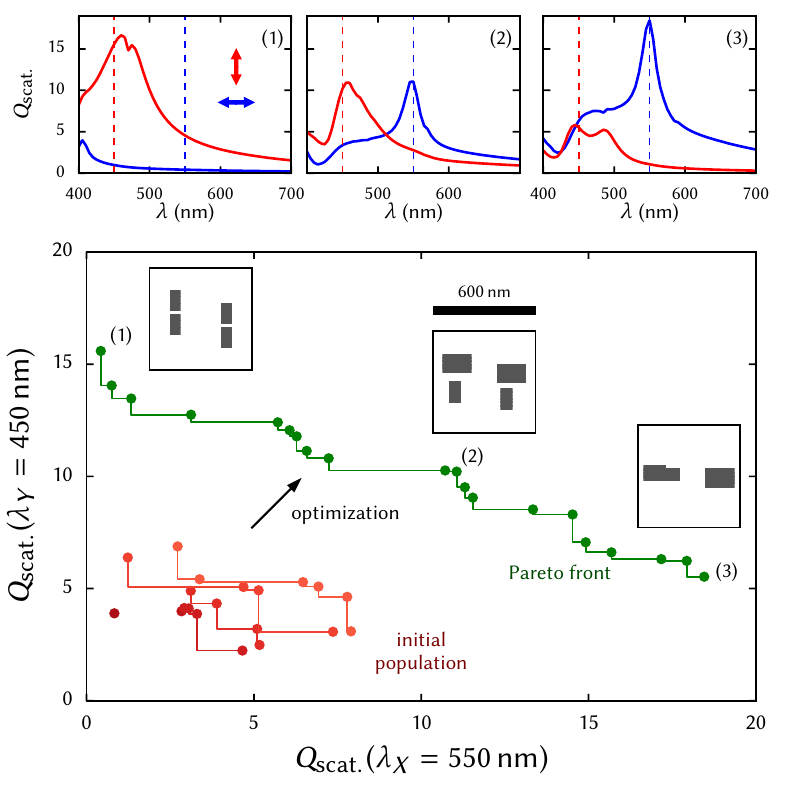} 
\caption{
\textbf{Pareto front example of an optimisation run with \(\lambda_X=550\,\)nm and \(\lambda_Y=450\,\)nm.}
Top: Spectra of selected antennas (indicated by numbers on Pareto front), where either a single wavelength is optimised (1 and 3) or both resonance wavelengths are scattered approximately equally (2). 
\(X\) (\(Y\)) polarised illumination is plotted with blue (red) colour. 
The selected structures are sketched in the insets, showing areas of \(600 \times 600\,\)nm$^{2}$.
}
\label{fig:pareto_450_550}
\end{figure}

In a next step we study the maximisation of \(Q_{\text{scat}}\) at two different wavelengths \(\lambda_X=550\,\)nm and \(\lambda_Y=450\,\)nm for mutually crossed polarisations.
The randomly initialised population of \(20\) individuals at the beginning of the evolution (red), the Pareto front (green) and selected structure designs as well as corresponding spectra are shown in figure~\ref{fig:pareto_450_550}.
The individuals at the Pareto front borders, labelled (1) and (3), correspond to single-objective optimisations for \(\lambda_Y\) and \(\lambda_X\), respectively. Inspecting the three selected structures in more detail leads to the following observations.

Obviously twin structures like (1) and (2) seem to be preferred, because they result in an increase of the overall scattering efficiency. 
Indeed, structures (1) and (2) both consist of two dimer antennas that, if taken individually, have about \(30\,\)\%, respectively \(10\,\)\%, lower \(Q_{\text{scat}}\) at the target wavelength of \(\lambda_Y=450\,\)nm compared to the twin structure. 
Furthermore, the peak positions in the scattering spectra are slightly shifted and match the target wavelengths only in the combined antenna.

We point out that the rather symmetric relative positioning of the two dimers is crucial for an optimum scattering efficiency.
The configuration found by the evolutionary optimisation is very close to the ideal positions. 
A marginally stronger scattering can be obtained for both structures (1) and (2), when the dimers are placed on the same horizontal axis but the possible gain is as low as about \(3\,\)\% and \(1\,\)\%, respectively.

At last, particle (3) in Fig.~\ref{fig:pareto_450_550} consists only of a single dimer structure, which we attribute to the constrained maximum antenna size in our model.
The maximisation of the scattering at the longer target wavelength (\(\lambda_X = 550\,\)nm) requires a larger amount of material compared to shorter wavelength \(\lambda_Y\). 
The scattering efficiency can be further improved by allowing the algorithm to use larger or more constituents.

In the supporting informations, a detailed analysis of structures~(1) and~(2) is shown ({\color{blue}section~E, Figs.~S6 and~S7}), as well as a demonstration that the different geometry of structure~(3) can be explained by the limited amount of silicon allowed in the computation ({\color{blue}section~F}).

\begin{figure*}[t]
\centering
\includegraphics[page=1]{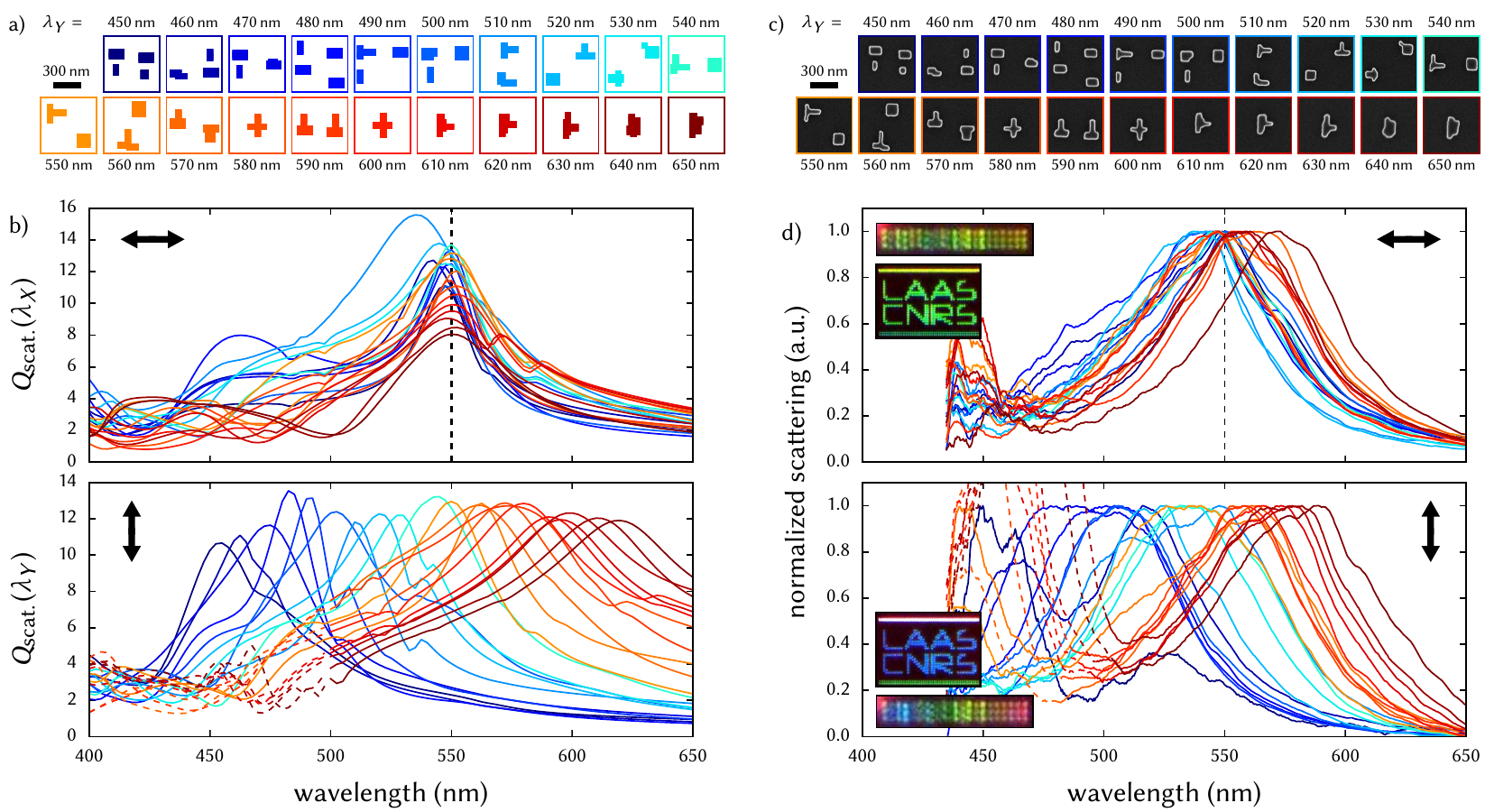}
\caption{
\textbf{Experimental demonstration of several dual-resonant silicon structures based on evolutionary multi-objective optimisation.}
(a) EMO design of multi-resonant dielectric particles and (b) simulated scattering spectra for \(\lambda_X=550\)\,nm (indicated by a black dashed line) and various \(\lambda_Y\). 
(c) SEM images and (d) polarisation filtered scattering spectra of the corresponding nanofabricated sample, normalized to the peak closest to the target wavelength. 
Insets in (d) show polarisation filtered darkfield microscopy images of the full set of structures (\(20\times 4\,\)\textmu m\(^2\)) and of the LAAS laboratory logo (\(35\times 23\,\)\textmu m\(^2\)). 
The lines framing the blue letters (\(\lambda_Y=450\,\)nm) are optimised for \(\lambda_Y=650\,\)nm (upper line) and \(\lambda_Y=570\,\)nm (lower line).
Areas in a) and c) are \(600 \times\ 600\,\)nm\(^2\).
}
\label{fig:tuning_structures}
\end{figure*}

To further illustrate the EMO-GDM technique, we perform several multi-objective optimisations for different combinations of target wavelengths.
The wavelength \(\lambda_X=550\,\)nm is fixed, while the other (\(\lambda_Y\)) is varied from \(450\,\)nm to \(650\,\)nm in steps of \(10\,\)nm. 
Each simulation consists of an initial population of 20 random individuals, which is evolved for 200 generations.
At the end of the evolution, the optimised structure having the closest scattering efficiencies for \(X\) and \(Y\) polarisation, \emph{i.e.}
\begin{equation}
 |Q_{\text{scat}}(\lambda_X) - Q_{\text{scat}}(\lambda_Y)| = \text{min.} 
\end{equation}
is chosen from each simulation (like structure (2) in Fig.~\ref{fig:pareto_450_550}).

In Figure~\ref{fig:tuning_structures}, we show the resulting structures (a) and their GDM-simulated spectra for \(X\)- and \(Y\)-polarised incidence (b).
The different \(\lambda_Y\) are indicated by a colour coding from blue (\(\lambda_Y=450\,\)nm) to red (\(\lambda_Y=650\,\)nm).
As explained in the previous subsection, for increasing wavelengths, the four sub-antennas tend to combine in only two structures (instead of more constituents for the shortest wavelengths), which is due to the limited amount of allowed material. 
For the same reason, at wavelengths above \(600\,\)nm all sub-antennas are even merged into one single structure, and for the longest wavelengths the available material is not sufficient to yield a satisfactory maximisation
(for an analysis of the role of the constrained amount of material, see {\color{blue} SI, Secs.~F and~G}).

For an experimental verification, we fabricated Si-structures corresponding to the optimised colour-tuned nanoantennas. 
SEM images (Fig.~\ref{fig:tuning_structures}c) and polarisation filtered darkfield spectra (Fig.~\ref{fig:tuning_structures}d, top: filter along \(X\), bottom: along \(Y\)) are shown in figure~\ref{fig:tuning_structures}. 
Polarisation filtered darkfield images (Fig.~\ref{fig:sketch_optimization}d and insets in Fig.~\ref{fig:tuning_structures}d) of colour-switching pictograms, composed of the optimised structures, demonstrate the polarisation dependence of the scattered wavelengths.
For the largest structures, a high order resonance appears in the blue part of the spectrum. 
This resonance, already visible in the simulations and experiments of Figure~\ref{fig:pareto_630_630}, is shown in Figure~\ref{fig:tuning_structures}b and~d with dashed lines to distinguish it from the low order resonance of the smaller nanostructures which appear in the same spectral region. 
Simulated and experimental data in Figure~\ref{fig:tuning_structures} have been normalized at the targeted wavelength.
In the experiment, the relative contribution of the high order resonance to scattering is reinforced by the Si/SiO\(_2\)/Si cavity of the underlying SOI substrate. 
We provide in {\color{blue}SI section~I} the complete set of experimental and computed data to allow for a quantitative comparison and discuss the influence of the underlying cavity for the largest nanostructures.

By a closer look on the individual structures, we observe that the ``symmetric'' optimisation with \(\lambda_X = \lambda_Y = 550\,\)nm results in a non-symmetric particle.
We would intuitively expect a symmetric antenna to be ideally suited for equally strong scattering under both, \(X\)- and \(Y\)-polarisation.
The evolutionary optimisation, being a non-analytic routine, should at least result in some ``quasi''-symmetric structures, which is however not the case here.
As before, this can be explained by the finite amount of material available in our structure model. 
Because the T-shaped part of the antenna already consists of three of the four sub-antennas, the fourth sub-antenna is added as a square block of maximum allowed dimension, and it is impossible for the algorithm to generate a symmetric structure within the given constraints.
As shown in the supporting informations, a simulation with \(\lambda_X = \lambda_Y = 450\,\)nm as well as an optimisation with relaxed constraints on the antenna size results in quasi-symmetric structures, as intuitively expected ({\color{blue}see SI, Sec.~H}).

Again, for \(\lambda_X = \lambda_Y = 550\,\)nm, interference between both parts of the antenna results in an optimum scattering efficiency at the target wavelength and therefore exact positioning of the constituents is crucial: 
A change of the spacing between the T-shaped and squared sub-structures by \(\Delta x = 100\,\)nm already results in a decrease of more than \(5\)\,\% in scattering efficiency for at least one polarisation.
An analysis of the \(\lambda_X = \lambda_Y = 550\,\)nm antenna can be found in the supporting informations ({\color{blue}Sec.~E, Fig.~S8}).

\subsection*{Polarisation encoded micro images}

\begin{figure}[tb]
\centering
\includegraphics[page=1]{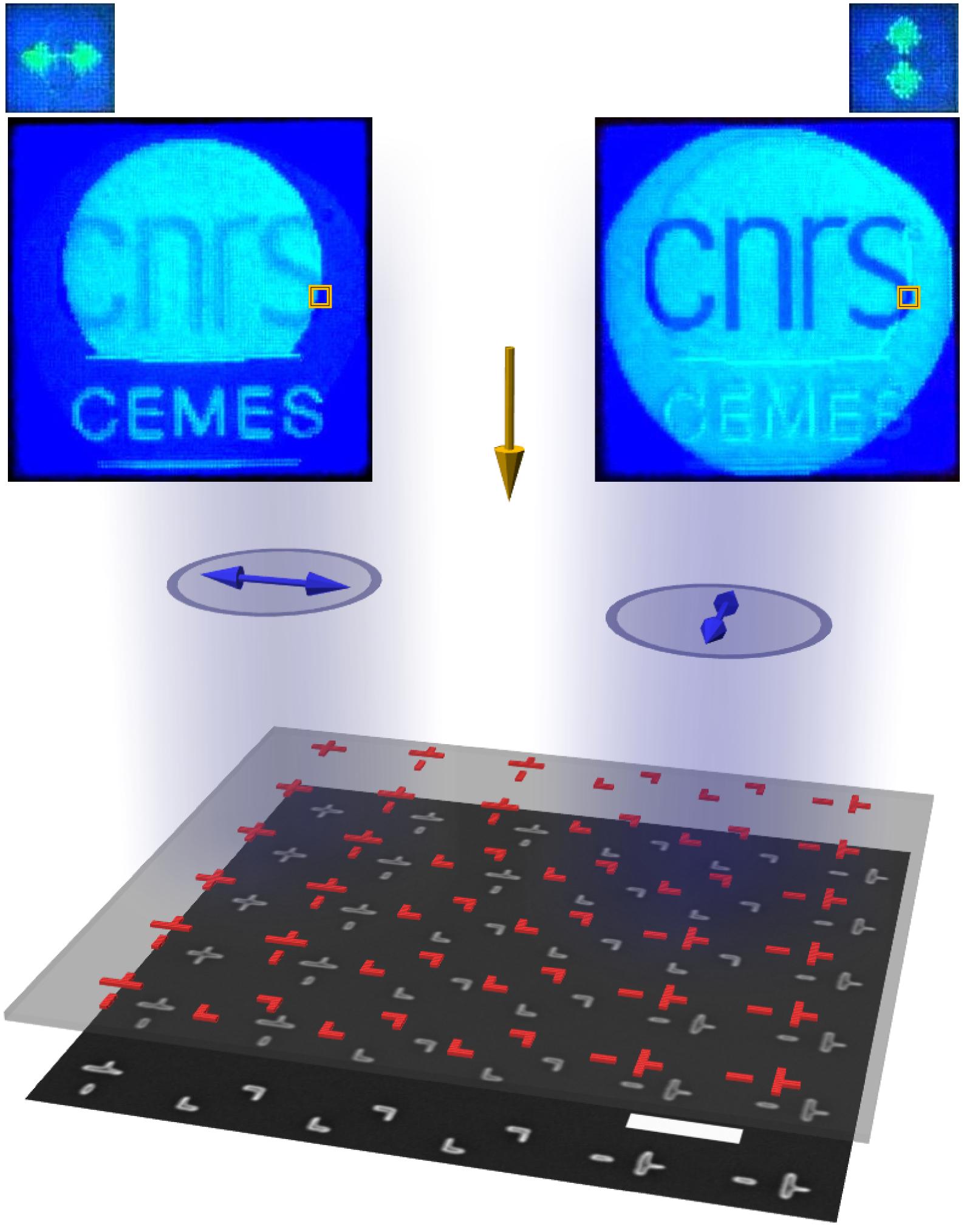} 
\caption{
\textbf{Polarisation-filtered darkfield images of micrometer scale pictures designed by evolutionary optimisation.}
Micrometre scale pictures composed of \(24 \times 24\) (arrows) and \(100 \times 100\) (logos) EMO-GDM designed particles.
A linear polarisation filter is added before the camera, oriented along \(X\) (top, left) and along \(Y\) (top, right). 
Arrow images are \(15 \times 15\,\text{\textmu m}^2\), logos \(60 \times 60\,\text{\textmu m}^2\) large.
Bottom image: Zoom into the logo-picture. SEM image in grey (scalebar is \(500\,\text{nm}\)) and sketch of the lithographic mask in red, highlighted by small yellow squares in the darkfield images.
The yellow arrow and blue emission indicate the incident and scattered light, respectively.
}
\label{fig:darkfield_images}
\end{figure}

To illustrate the previous results we produced small images, only few micrometres large, composed of EMO-optimised antennas. 
The absolute scattering cross section \(\sigma_{\text{scat}}\) was used as the optimisation target. 
An additional spacing of \(250\,\)nm is used between the individual particles, which results in pixel sizes of \(850 \times 850\,\)nm\(^2\) (\(\approx 30000\,\)dpi), close to the diffraction limit.

Polarisation-filtered dark field images are shown in figure~\ref{fig:darkfield_images}.
Depending on the orientation of the polarisation filter (left: \(X\), right: \(Y\)), one single arrow is visible, pointing in the corresponding direction while the second arrow vanishes in a blue background.
Furthermore, the logos of the CNRS and CEMES laboratory are nested into one image, encoded in perpendicular polarisations. 
A scheme of the lithographic mask (red) and a SEM image (grey) of a zoom into the logos, indicated by small yellow squares, is shown at the bottom. 
We attribute the slightly reminiscent signatures of the hidden motifs to intensity-variations due to the arrangement of the antennas in grating-like 2D-arrays ({\color{blue}see also SI Sec.~J}).

\subsection*{Conclusions}

In conclusion, we presented a technique of evolutionary multi-objective optimisation coupled to full-field electro-dynamical simulations for the automatic design of photonic nanostructures.
We demonstrated that our approach is able to design double-resonant silicon nanoantennas even within a very simple structure-model. 
We found that all accessible parameters were nearly perfectly optimised by the evolutionary algorithm.
Furthermore, for a maximum compatibility with fabrication methods, technological limitations were included as boundary conditions in the model.
Thanks to these additional requirements, the measured spectra of samples produced on SOI substrate showed an excellent agreement with the predictions of the optimisations.

A great advantage of the EMO-GDM technique is its flexibility and the ability to self-adapt to arbitrary limitations.
Additional constraints can easily be implemented because no analytical treatment of the input model needs to be performed.
Inadequate structures, inconsistent with the constraint functions, are being discarded automatically during the evolution and only technologically convenient designs are generated.
The method can also be easily extended for the rigorous design of metasurfaces, where interference between the unit cells needs to be considered.
Periodic boundary conditions can be included in the GDM by means of an appropriate Green's Dyad.\cite{chaumet_simulation_2009, gallinet_electromagnetic_2009}
In this way, the distance between substructures on the metasurface may also be included as a free parameter in the optimisation.
We believe that multi-objective optimisation of photonic nanostructures has a tremendous potential for many kinds of possible applications in near- and far-field nano-optics for example in the design of multiresonant, broadband light harvesting, or nonlinear nanostructures.

\section*{Methods}

\subsection*{EMO-GDM method}
We use the python interface of the parallel evolutionary multi-objective optimisation (EMO) toolkit paGMO/pyGMO\cite{biscani_global_2010} and in particular its implementation of the ``SMS-EMOA'' algorithm\cite{beume_sms-emoa:_2007}. 
A comprehensive introduction to evolutionary multi-objective optimisation can be found in reference~\onlinecite{deb_multi-objective_2001}.

All interfacing between the EMO and the electro-dynamical full-field solver is implemented in python.
The fitness of each nanoparticle is calculated using the Green Dyadic Method (GDM), which is implemented in fortran to yield high computational speed.

The target nanoparticle is discretised in \(N\) cubic meshpoints of side-length \(b\), for each of which a dipolar response is assumed. 
This approach eventually leads to a system of \(3N\) coupled equations that relates an incident electric field \(\mathbf{E}_0\) to the field \(\mathbf{E}\) due to the particle's response:
\begin{equation}
 \mathbf{E}_0 = \mathbf{M} \cdot \mathbf{E}.
\end{equation}
The field in the structure can then be obtained by an inversion of the matrix \(\mathbf{M}\), which is composed of \(3\times 3\) sub-matrices 
\begin{equation}
 \mathbf{M}_{ij} = \mathbf{I}\, \delta_{ij} - \alpha_i(\omega)\, \mathbf{G}(\mathbf{r}_i, \mathbf{r}_j, \omega).
\end{equation}
Here, \(\mathbf{I}\) is the Cartesian unitary tensor, \(\delta_{ij}\) the Kronecker delta function and (cgs units)
\begin{equation}
 \alpha_i(\omega) = \frac{\epsilon_{i}(\omega) - \epsilon_{\text{env}}(\omega)}{4\pi} v_i,
\end{equation}
\(v_i\) is the volume of each cubic cell, in our case \(v_i = b^3\). 
For the permittivity \(\epsilon_{i}\) we use the dispersion of silicon from Ref.~\onlinecite{palik_silicon_1997} and assume a constant environment of \(\epsilon_{\text{env}}=1\).

\(\mathbf{G}\) is the Green's Dyad which couples the dipolar elements \(i\) and \(j\) and is composed of a vacuum and a surface term
\begin{equation}\label{eq:GreensDyadFull}
 \mathbf{G}(\mathbf{r}_i, \mathbf{r}_j, \omega) = 
  \mathbf{G}_{0}(\mathbf{r}_i, \mathbf{r}_j, \omega)
    +
  \mathbf{G}_{\text{surf}}(\mathbf{r}_i, \mathbf{r}_j, \omega)
\end{equation}
which can be found in literature.\cite{girard_shaping_2008}
To account for the divergence of the Green's function at \(\mathbf{r}_i=\mathbf{r}_j\), a normalization scheme 
\begin{equation}
 \mathbf{G}_{0}(\mathbf{r}_i, \mathbf{r}_i, \omega) = \mathbf{I}\, C(\omega)
\end{equation}
is introduced, which writes for a cubic mesh
\begin{equation}\label{eq:normalizationTerm}
 C(\omega) = -\frac{4\pi}{3} \frac{1}{\epsilon_{\text{env}}(\omega) v_i}
\end{equation}
and has to be adapted together with the cell volume, if a different meshing is used like, for example a hexagonal compact grid.\cite{girard_shaping_2008}
Note that we neglected a weak radiative term in Eq.~\eqref{eq:normalizationTerm}, which is discussed in Ref.~\onlinecite{girard_shaping_2008} and references therein.

Finally, the matrix inversion is done using standard LU-decomposition and
the scattering efficiencies can be calculated from the near-field \(\mathbf{E}\) inside the particle.\cite{draine_discrete-dipole_1988}

A great advantage of the GDM is that the presence of a substrate (in our case \(n=1.5\)) can be taken into account by means of an appropriate Green's Dyadic function (Eq.~\eqref{eq:GreensDyadFull}), which can be calculated in the non-retarded approximation at almost no supplementary computational cost.
In comparison to finite-difference time-domain (FDTD) simulations, a frequency-domain method has further advantages with regards to our purpose of designing a doubly resonant nanostructure:
So-called perfectly matched layers are not needed and only the nanoparticle itself is subject to the volume discretisation. Generally, this results in a quicker convergence.

The amount of silicon per antenna is not constant in our model and as the duration of a simulation is depending on the structure size, the optimisations generally tend to be slower for longer resonance wavelengths, because of resulting larger particles.
Nevertheless, evolutions of populations with 20 individuals over 200 generations take not longer than around 10-15 hours on one single core of a \(2.8\)\,GHz Intel Xeon E5-1603 CPU.
We note that the results were always reproducible, yielding very similar structures and scattering efficiencies from multiple runs ({\color{blue} see SI, section~D}).

\subsection*{Nanoantenna fabrication by top-down approach}
Samples were fabricated by a top-down approach that couples Electron Beam Lithography (EBL) with anisotropic plasma etching. 
This was used to pattern the designed nanostructures \cite{han_realization_2011, guerfi_high_2013} on a silicon-on-insulator (SOI) wafer as substrate (Si: 95 nm, BOX: 145 nm). 
The EBL was carried out with a RAITH 150 writer at an energy of 30 keV on a thin (60nm) negative-tone resist layer, namely hydrogen silsesquioxane (HSQ). 
After exposure, HSQ was developed by immersion in \(25\,\)\% tetramethylammonium hydroxide (TMAH) for \(1\,\)min. 
HSQ patterns were subsequently transferred to the silicon top layer by reactive ion etching in a SF6/C4F8 plasma based chemistry down to the buried oxide layer.

In the EMO runs, the minimum feature size was set to \(60\,\)nm to avoid removing small features of the structures during lift-off.  
The structures were discretised and placed on a grid by steps of \(20\,\)nm to match the precision of the EBL.
SEM images of individual structures are shown and are compared to the mask-layout in Figs.~\ref{fig:pareto_630_630}, \ref{fig:tuning_structures} and \ref{fig:darkfield_images}.

\subsection*{Confocal darkfield microscopy}
Confocal optical darkfield microscopy was performed on a conventional spectrometer (Horiba XploRA).
A spectrally broad white lamp was focused on the sample by a \(\times 50\), NA\,\(0.45\) darkfield objective, backscattered, polarisation filtered and dispersed by a \(300\,\)grooves per mm grating onto an Andor iDus 401 CCD.
The intensity distribution of the lamp as well as the spectral response of the optical components was accounted for by subtracting the background measured on bare SOI and normalizing the measured spectra to a white reference sample.


\begin{acknowledgments}
The authors thank P. Salles and G.-M. Caruso for technical assistance. 
This work was partly supported under ``Campus Gaston Dupouy'' grant by French government, R\'egion Midi-Pyr\'en\'ees and European Union (ERDF), by the computing facility center CALMIP of the University of Toulouse under grant P12167 and by LAAS-CNRS micro and nanotechnologies platform member of the French RENATECH network.
\end{acknowledgments}

\section*{Author contributions}
P.R.W., V.P. and A.A. designed the research. 
C.G., A.A. and P.R.W. implemented the codes and performed the simulations. 
A.L. and G.L. fabricated the samples by EBL. 
P.R.W and V.P. performed the darkfield scattering experiments. 
All authors contributed to the data analysis, figure preparation and manuscript writing.

\section*{Additional information}
Supplementary information is available in the online version of the paper. Reprints and permission information is available online at www.nature.com/reprints. Correspondence and requests for materials should be addressed to P.R.W., A.A. and V.P.

\section*{Competing financial interests}
The authors declare no competing financial interests.

\newpage
\widetext
%
%
 \begin{center}
\section*{Supporting Informations: \TITLE}
 \end{center}
\setcounter{equation}{0}
\setcounter{figure}{0}
\setcounter{table}{0}
\setcounter{page}{1}
\makeatletter
\renewcommand{\theequation}{S.\arabic{equation}}
\renewcommand{\thefigure}{\arabic{figure}}
\renewcommand{\bibnumfmt}[1]{[S#1]}
\renewcommand{\citenumfont}[1]{S#1}
%
%



\subsection{Number of possible parameter permutations in the structure model}

\begin{figure}[h!]
\centering
\includegraphics[page=1]{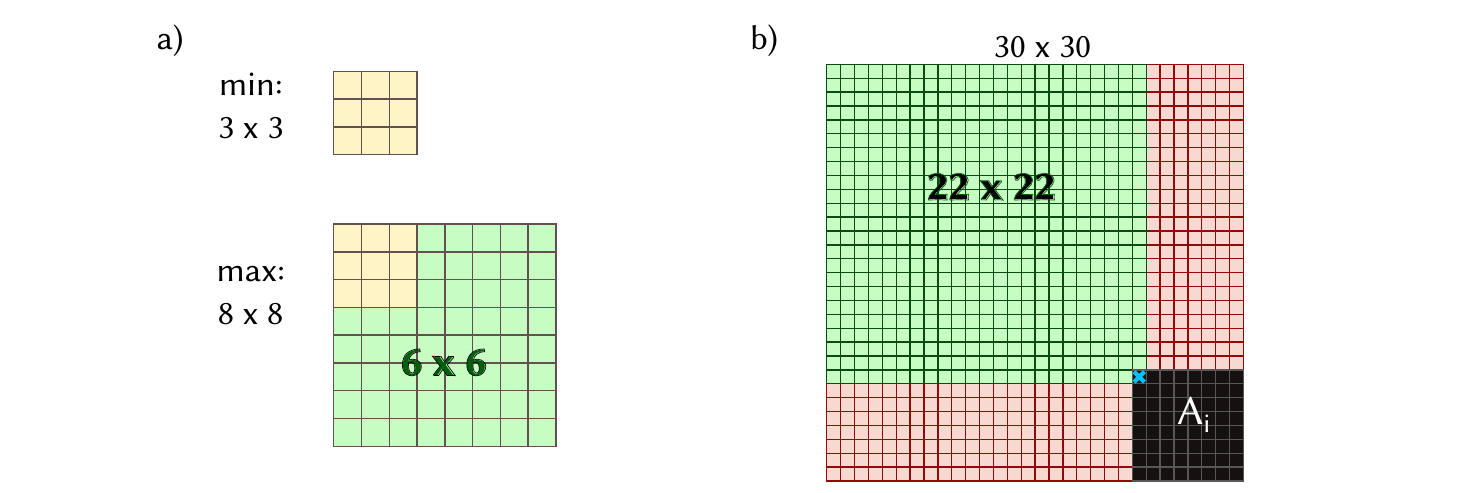} 
\caption{\FIGCAPTIONPREFIX \textbf{Scheme for estimating the total number of possible parameter permutations in the structure model.} a) Scheme illustrating the number of possible parameter permutations resulting from the size-variations of each sub-block. b) Scheme illustrating the number of possible parameter permutations resulting from the positioning of each sub-block on the computationally considered grid.}
\label{figSI:permutations}
\end{figure}

We can estimate the number of possible parameter combinations:
Each block can be varied in size between \(3 \times 3\) and \(8 \times 8\) units of \(20\,\)nm, resulting in \(6 \times 6\) possible permutations per sub-antenna (see Fig.~\ref{figSI:permutations}a).
Four of those antennas are then placed on a field of \(600\,\text{nm} \times 600\,\text{nm}\) which is divided by steps of \(20\,\)nm.
This gives a total of \(22 \times 22\) positions (Fig.~\ref{figSI:permutations}b), assuming a \(8 \times 8\) block, which is a lower limit for the actual possible number of positions.
We get
\begin{equation}
 \begin{aligned}
 & \binom{(22 \times 22) \times (6 \times 6)}{4} \\
 & \gtrsim 1 \times 10^{15}.
 \end{aligned}
\end{equation}
Identical arrangements are considered by the permutation operator \(\binom{a}{b}\) (``b out of a''), but identical structures due to symmetry are not taken into account. 

\clearpage
\subsection{Stepsize in Green Dyadic simulations}

\begin{figure}[h!]
\centering
\includegraphics[page=1]{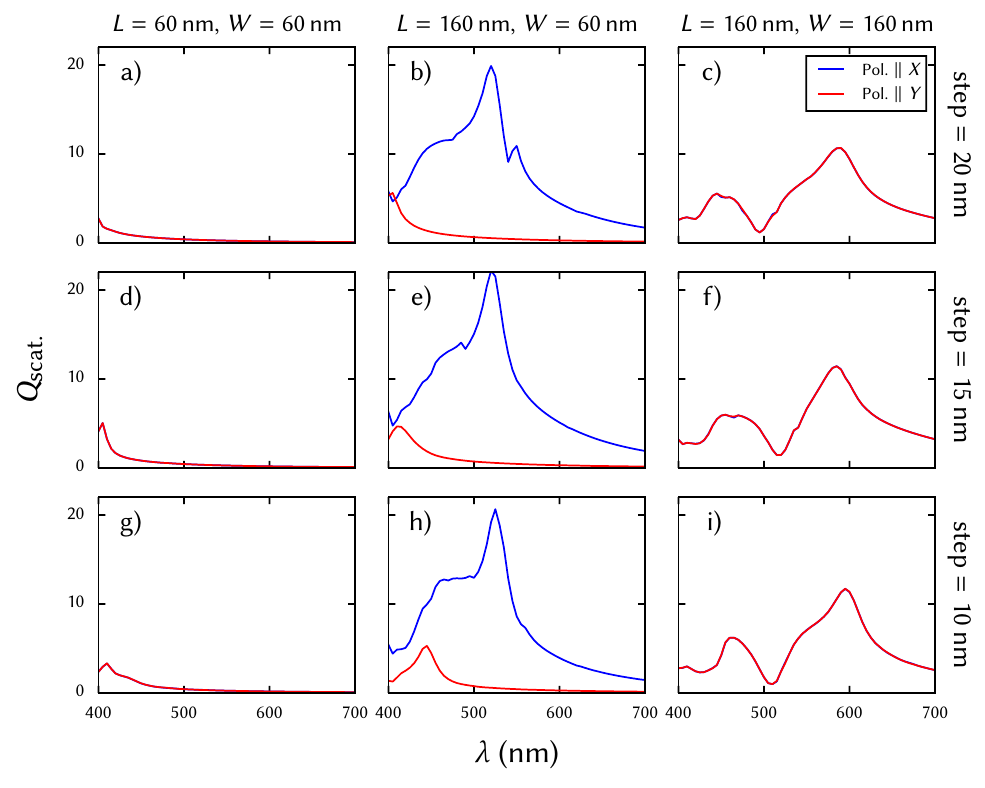} 
\caption{\FIGCAPTIONPREFIX \textbf{Simulations of identical structures with different discretisation stepsizes.} Spectra for cuboidal silicon blocks of height \(H=100\,\)nm and width / length combinations corresponding to the minimal and maximal allowed dimensions. Simulations were performed with different discretisation stepsizes \(S=20\,\)nm (a-c), \(S=15\,\)nm (d-f) and \(S=10\,\)nm (g-i).}
\label{figSI:stepsize}
\end{figure}

To verify that a large discretisation of \(20\,\)nm can be used in the GDM simulations to yield correct scattering efficiencies, we performed simulations of identical geometries with different discretisation stepsizes.
Fig.~\ref{figSI:stepsize} shows spectra obtained using stepsizes of \(S=20\,\)nm (top), \(S=15\,\)nm (center) and \(S=10\,\)nm (bottom). 
In the structure with \(S=15\,\)nm, \(H=105\,\)nm (and if applicable \(L,W=165\,\)nm) were used instead of \(H=100\,\)nm (and \(L,W=160\,\)nm).
The good qualitative and quantitative agreement justifies the use of a stepsize of \(20\,\)nm in order to speed up the optimisation process.

\subsection{Structures in initial population vs. Pareto front for \(\lambda_X = \lambda_Y = 630\,\)nm}

\begin{figure}[h!]
\centering
\includegraphics[page=1]{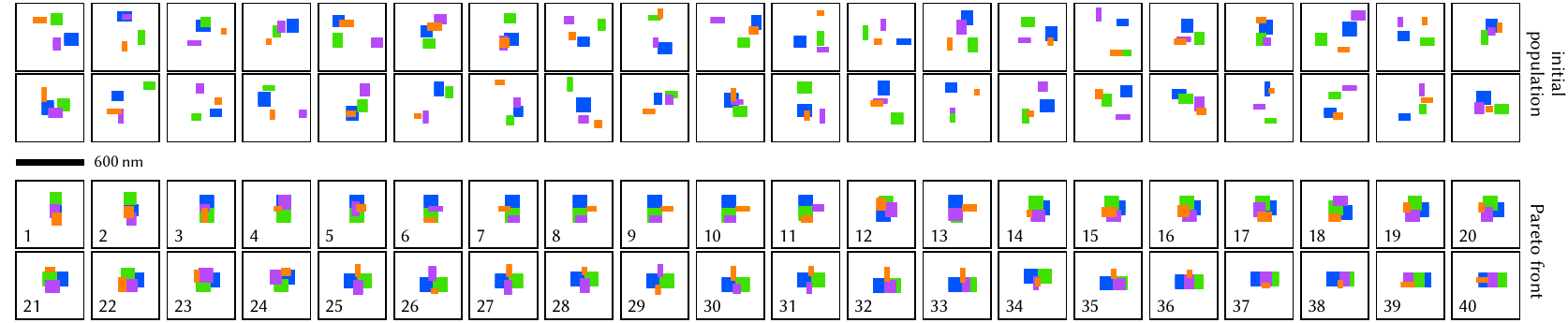} 
\caption{\FIGCAPTIONPREFIX \textbf{Structure population before and after evolutionary multi-objective optimisation.} Structures of the initial population (two top rows) vs. structures on the Pareto-front after the EMO (two bottom rows). 
Individual sub-blocks of the structure model are highlighted with different colours. 
Data corresponding to main paper, figure~3.}
\label{figSI:pareto630630structures}
\end{figure}

\clearpage
\subsection{Convergence of optimisations / reproducibility}

Assessing the convergence of evolutionary algorithms to a global optimum and the reproducibility of their predictions from different simulation runs are two central issues.
In contrast to evolutionary approaches, classical steepest-descent techniques, \emph{e.g.} variations of the Newton-Raphson method, converge rapidly to local extrema. 
While steepest-descent type methods adapted to multi-objective problems exist,\cite{fliege_steepest_2000} their fast convergence towards local extrema often limits their reliability for the determination of global maxima.
A possible solution are hybrid algorithms, combining steepest-descent type techniques with evolutionary procedures, which are subject of current algorithm development.\cite{giacomini_comparison_2014, desideri_multiple-gradient_2014, chen_new_2015}
However, in our case such algorithms are not applicable since steepest-descent requires continuously differentiable target functions (see eg. Ref.~\onlinecite{fliege_steepest_2000}). 
Our target, represented by the scattering of a discrete nanostructure placed on a discretised grid, is neither differentiable nor continuous.
Hence we are constrained to purely heuristic algorithms.

\begin{figure}[h!]
\centering
\includegraphics[page=1]{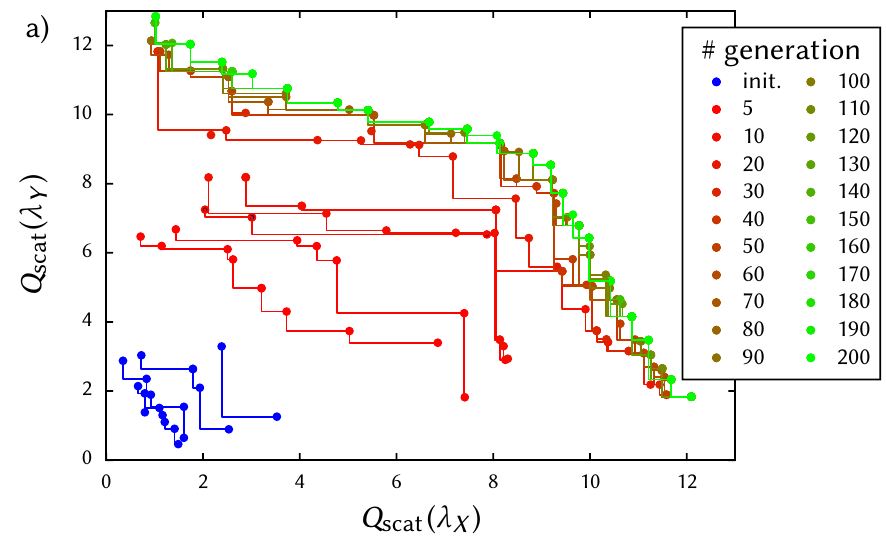} 
\hspace{.5cm}
\includegraphics[page=1]{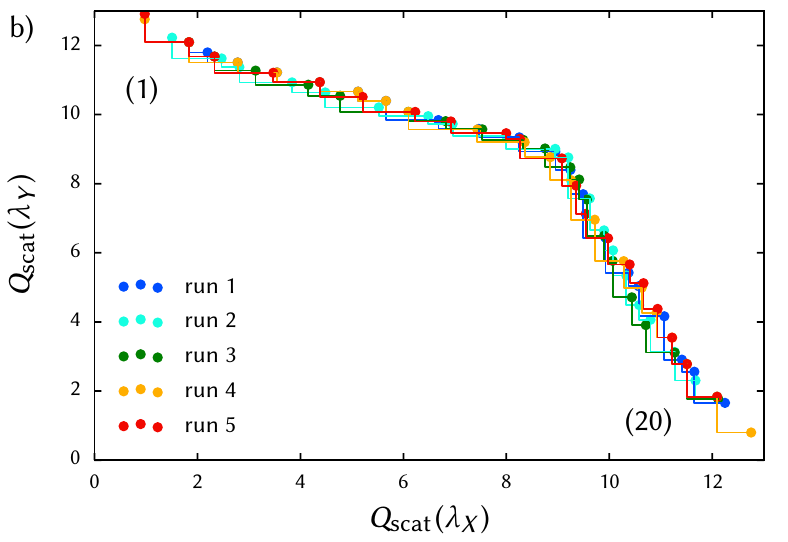} 
\caption{\FIGCAPTIONPREFIX 
\textbf{Convergence and reproducibility of the evolutionary multi-objective optimisation.}
a) Evolution of the population during an optimisation with \(\lambda_X = \lambda_Y = 630\,\)nm, showing convergence after around \(100\) iterations.
b) Pareto-fronts of several independent EMO-GDM runs each with random initial population of 20~individuals, but identical objective functions.
}
\label{figSI:repeatability_630630_paretofront}
\end{figure}

In order to verify that the evolutionary optimisation converges, we plot the Pareto-front at each \(10\)th iteration during the evolution, which is shown in Fig.~\ref{figSI:repeatability_630630_paretofront}a. 
The optimisation converges after around \(100\) iterations, when the Pareto-front stops expanding.
The shown results are for \(\lambda_X = \lambda_Y = 630\,\)nm, equal to the case shown in the main paper, figure~3.

\begin{figure}[h!]
\centering
\includegraphics[page=1]{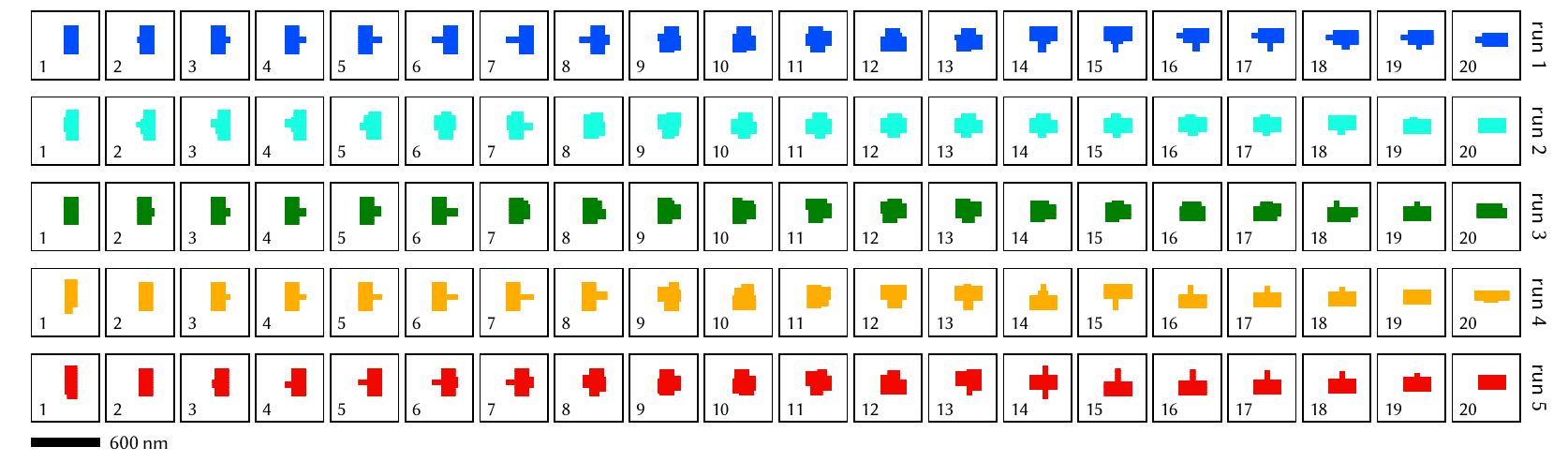} 
\caption{\FIGCAPTIONPREFIX 
\textbf{Comparison of final populations from several optimiation runs.}
Structures corresponding to the Pareto-fronts shown in Fig.~\ref{figSI:repeatability_630630_paretofront}b (same colour coding), sorted from lowest to highest \(Q_{\text{scat.}}(\lambda_X)\). 
Areas are \(600 \times 600\,\text{nm}^2\) large.}
\label{figSI:repeatability_630630_structs}
\end{figure}

To check the reproducibility of the optimisations, we performed several successive runs of the same problem (\(\lambda_X = \lambda_Y = 630\,\)nm).
The Pareto-fronts and the corresponding structures are plotted for five independent EMO-GDM evolutions of different randomly initialised populations of 20~individuals over 200 iterations in Fig.~\ref{figSI:repeatability_630630_paretofront}b, respectively Fig.~\ref{figSI:repeatability_630630_structs}, confirming the very good reproducibility of the optimisation. 
Both, the scattering efficiencies and the structure-layouts are very similar in all the different optimisation runs.
Finally we note that the results are also in agreement with the data presented in the main paper, obtained for identical target wavelengths but with a larger population of 40~individuals.

\enlargethispage{1cm}

\clearpage
\subsection{Analysis of optimised structures}

\begin{figure}[h!]
\centering
\includegraphics[page=1]{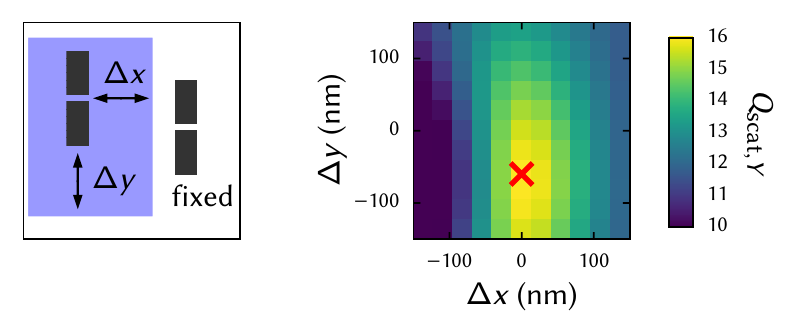} 
\caption{\FIGCAPTIONPREFIX 
\textbf{Analysis of relative positions in structure of optimisation with a single target \(\mathbf{\lambda_Y=450}\,\)nm.}
Left: Structure (1) from main paper Fig.~4. Blue highlighted part is shifted in \(x\)- and \(y\)-direction, shown area is \(600 \times 600\,\text{nm}^2\) large. 
Right: Scattering efficiency at \(\lambda_Y=450\,\)nm for polarisation along \(Y\) as function of displacement. The maximum is indicated by a red cross.}
\label{figSI:antenna_450_shift_analysis}
\end{figure}

\begin{figure}[h!]
\centering
\includegraphics[page=1]{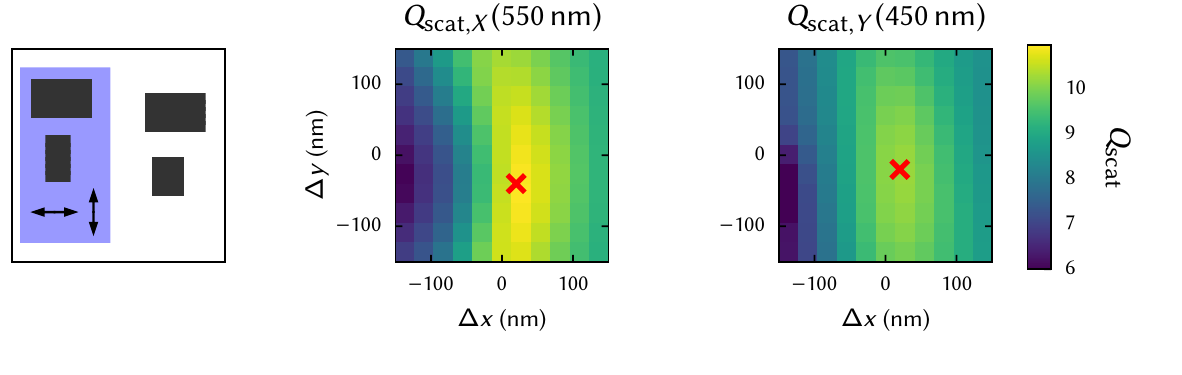} 
\caption{\FIGCAPTIONPREFIX 
\textbf{Analysis of relative positions in structure of optimisation with two targets \(\mathbf{\lambda_X = 550}\,\)nm and \(\mathbf{\lambda_Y = 450}\,\)nm.}
Left: Structure (2) from main paper Fig.~4 (\(\lambda_X = 550\,\)nm, \(\lambda_Y = 450\,\)nm, structure with most similar scattering). Blue highlighted part is shifted in \(x\)- and \(y\)-direction, shown area is \(600 \times 600\,\text{nm}^2\) large. 
Center, right: Scattering efficiencies at the target wavelengths for polarisation along \(X\) and \(Y\), respectively, as function of displacement of the sub-structure. The maximum in each colourplot is indicated by a red cross.}
\label{figSI:antenna_450550_shift_analysis}
\end{figure}

\begin{figure}[h!]
\centering
\includegraphics[page=1]{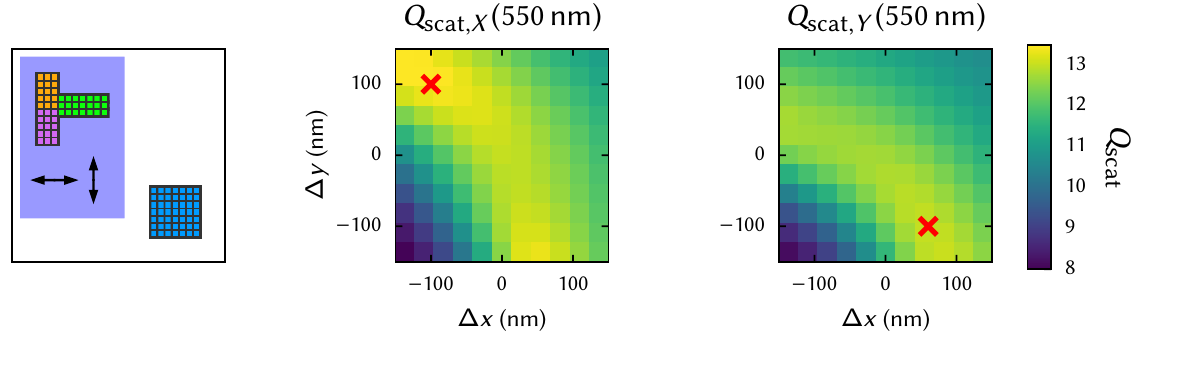} 
\caption{\FIGCAPTIONPREFIX 
\textbf{Analysis of relative positions in structure of optimisation with two identical targets \(\mathbf{\lambda_X = \lambda_Y = 550}\,\)nm.}
Left: Structure \(\lambda_X=\lambda_Y=550\,\)nm (see main paper Fig.~5). 
The sub-blocks are indicated by blue, green, purple and orange points.
The blue highlighted part is shifted in \(x\)- and \(y\)-direction, shown area is \(600 \times 600\,\text{nm}^2\) large.
Center, right: Scattering efficiencies at the target wavelengths for polarisation along \(X\) and \(Y\), respectively, as function of displacement of the sub-structure. The maximum in each colourplot is indicated by a red cross.}
\label{figSI:antenna_550550_shift_analysis}
\end{figure}

The positions of the constituents of antenna (1), shown in the main paper in Fig.~4 (single objective, maximum scattering for \(\lambda_Y=450\,\)nm) are almost ideally optimised. 
By a relative shift, an increase of only \(\approx 3\)\,\% in scattering efficiency is possible. 
Similarly, only a 1\,\% increase in scattering can be obtained for structure (2) of the same main-paper figure. 
In both cases, in an ideal configuration both sub-structures are on a horizontal mutual axis (see Figs.~\ref{figSI:antenna_450_shift_analysis} and~\ref{figSI:antenna_450550_shift_analysis}).
Considering again only the positions of the indicated components, the structure with \(\lambda_X=\lambda_Y=550\,\)nm from the main paper Fig.~5 actually reflects even the ideal configuration for a mutual optimisation of both polarisations (see Fig.~\ref{figSI:antenna_550550_shift_analysis}).
In all cases, a zero-shift corresponds to the positioning as found by EMO.

\enlargethispage{1cm}

\clearpage
\subsection{Larger allowed size: Single objective \(\lambda_{Y, \text{S.O.}} = 550\,\)nm and \(\lambda_{Y, \text{S.O.}} = 650\,\)nm}

We want to show that also for longer wavelengths several individual antennas are used to maximise scattering using interference. 
Therefore, single objective optimisations (individual of Pareto-front border) for \(\lambda_{Y, \text{S.O.}}=550\,\)nm and \(\lambda_{Y, \text{S.O.}}=650\,\)nm were ran, with a larger limit for the maximum available material (maximum side-length of \(300\,\)nm per block).
As expected, the resulting antennas are composed of several separated parts, each of which individually having a resonance at the target wavelength.

We conclude that the occurring difference in particle layout for increasing wavelengths is indeed a result of the limited amount of material available to the evolutionary algorithm.

The spike around \(470\,\)nm in the spectrum of the \(\lambda_{Y, \text{S.O.}}=650\,\)nm antenna is attributed to the large discretisation stepsize (see also below).

\begin{figure}[h!]
\centering
\includegraphics[page=1]{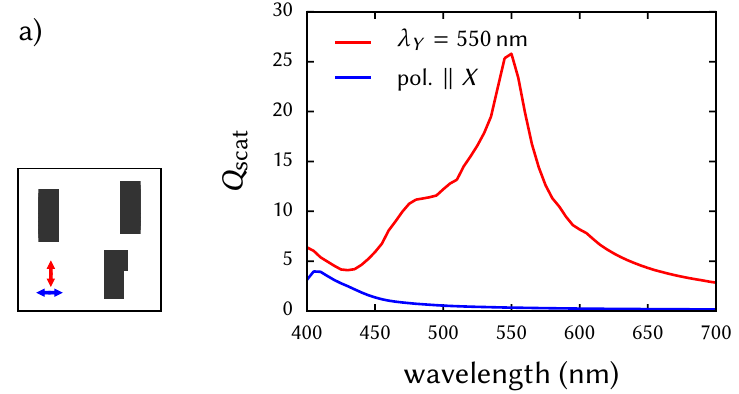} 
\hspace{1cm}
\includegraphics[page=1]{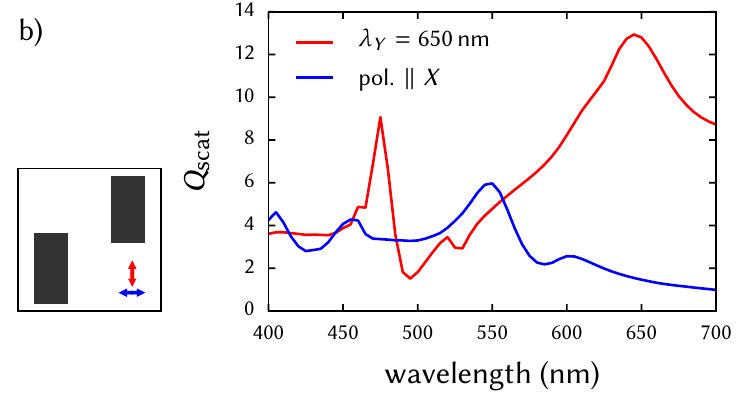} 
\caption{\FIGCAPTIONPREFIX 
\textbf{Optimisations with a single objective, allowing larger structures.}
Single-objective optimisation results for (a) \(\lambda_Y = 550\,\)nm and (b) \(\lambda_Y = 650\,\)nm with an incident polarisation along \(Y\) and using extended size constraints of \(L_i,W_i \in [40\,\text{nm}, 300\,\text{nm}]\) for each of the four blocks. 
On the left sketches of the structures are shown (showing \(700 \times 700\,\text{nm}^2\)).}
\label{figSI:larger_ant_SO550_SO650}
\end{figure}

\subsection{Larger allowed size: \(\lambda_X = 550\,\)nm and \(\lambda_Y=650\,\)nm}

\begin{figure}[h!]
\centering
\includegraphics[page=1]{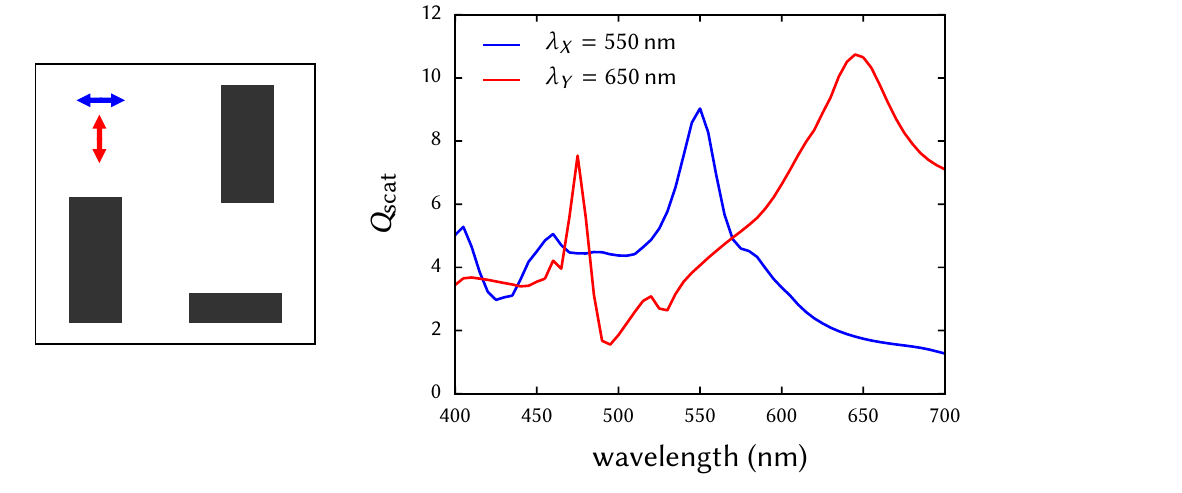} 
\caption{\FIGCAPTIONPREFIX 
\textbf{Optimisation with two objectives, allowing larger structures.}
EMO with \(\lambda_X = 550\,\)nm and \(\lambda_Y=650\,\)nm and extended size constraints of \(L_i,W_i \in [40\,\text{nm}, 300\,\text{nm}]\) for each of the four blocks. 
Left: Structure sketch (Frame shows \(700 \times 700\,\text{nm}^2\)). Right: Spectra for \(X\)- (blue) and \(Y\)-polarisation (red).}
\label{figSI:larger_ant_550_650_spectrum}
\end{figure}

The spectra of a simulation with large size-limits (maximum side-length of each block set to \(300\,\)nm) show, that a peak at \(\lambda_Y=650\,\)nm can be obtained if the optimisation algorithm is allowed to use more material (see Fig.~\ref{figSI:larger_ant_550_650_spectrum}).

The spike around \(470\,\)nm under \(Y\)-polarisation is attributed to the large discretisation stepsize, which can lead to artifacts in the spectral region of higher-order modes.

\clearpage
\subsection{Trial structures with \(\lambda_X = \lambda_Y\)}

If very small and very large features are allowed (\textit{i.e.} sufficient material is available to the EMO-algorithm), the optimisation of a structure with identical wavelengths for both polarisations results in (quasi-) symmetric structures.
Size limits are set to \(40\,\)nm for the lower and \(300\,\)nm for the upper boundaries.

\begin{figure}[h!]
\centering
\includegraphics[page=1]{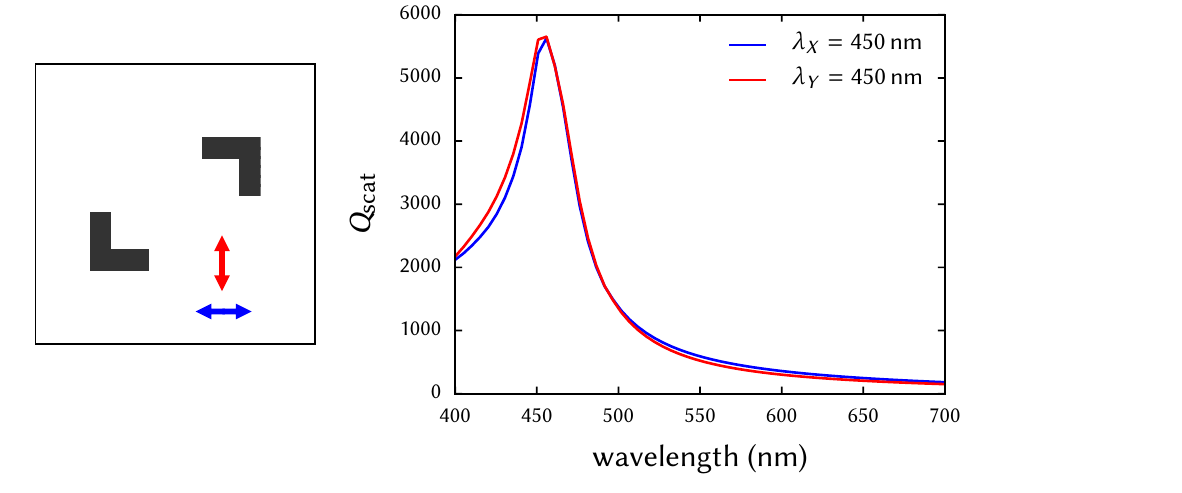} 
\caption{\FIGCAPTIONPREFIX 
\textbf{Less constrained optimisation with two identical objectives \(\mathbf{\lambda_X = \lambda_Y = 450}\,\)nm.}
EMO with \(\lambda_X = \lambda_Y = 450\,\)nm and extended size constraints of \(L_i,W_i \in [40\,\text{nm}, 300\,\text{nm}]\) for each of the four blocks. 
Left: Structure sketch (Area: \(600\,\text{nm} \times 600\,\text{nm}\)). Right: Spectra for \(X\)- (blue) and \(Y\)-polarisation (red).}
\label{figSI:antenna_450450_extended_size}
\end{figure}

\begin{figure}[h!]
\centering
\includegraphics[page=1]{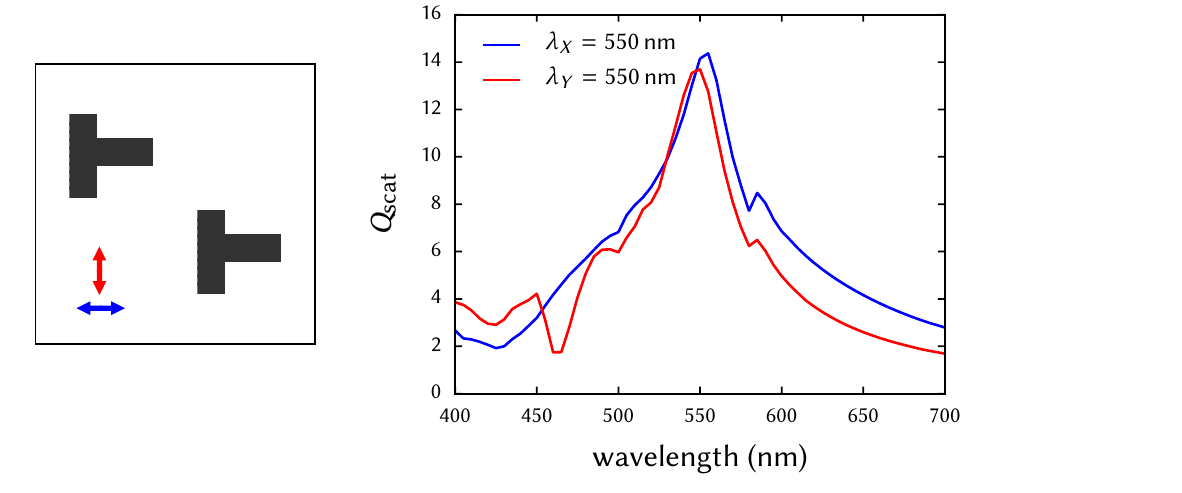} 
\caption{\FIGCAPTIONPREFIX
\textbf{Less constrained optimisation with two identical objectives \(\mathbf{\lambda_X = \lambda_Y = 550}\,\)nm.}
EMO with \(\lambda_X = \lambda_Y = 550\,\)nm and extended size constraints of \(L_i,W_i \in [40\,\text{nm}, 300\,\text{nm}]\) for each of the four blocks. 
Left: Structure sketch (Area: \(600\,\text{nm} \times 600\,\text{nm}\)). Right: Spectra for \(X\)- (blue) and \(Y\)-polarisation (red).}
\label{figSI:antenna_550550_extended_size}
\end{figure}

In Fig.~\ref{figSI:antenna_450450_extended_size} \(\lambda_X = \lambda_Y = 450\,\)nm is shown as an example.
Fig.~\ref{figSI:antenna_550550_extended_size} shows a simulation where \(\lambda_X = \lambda_Y = 550\,\)nm was used.

\clearpage
\subsection{Comparison of spectra for all double-resonant structures (main paper, Fig. 5)}

\begin{figure}[h!]
\centering
\includegraphics[width=\textwidth, page=1]{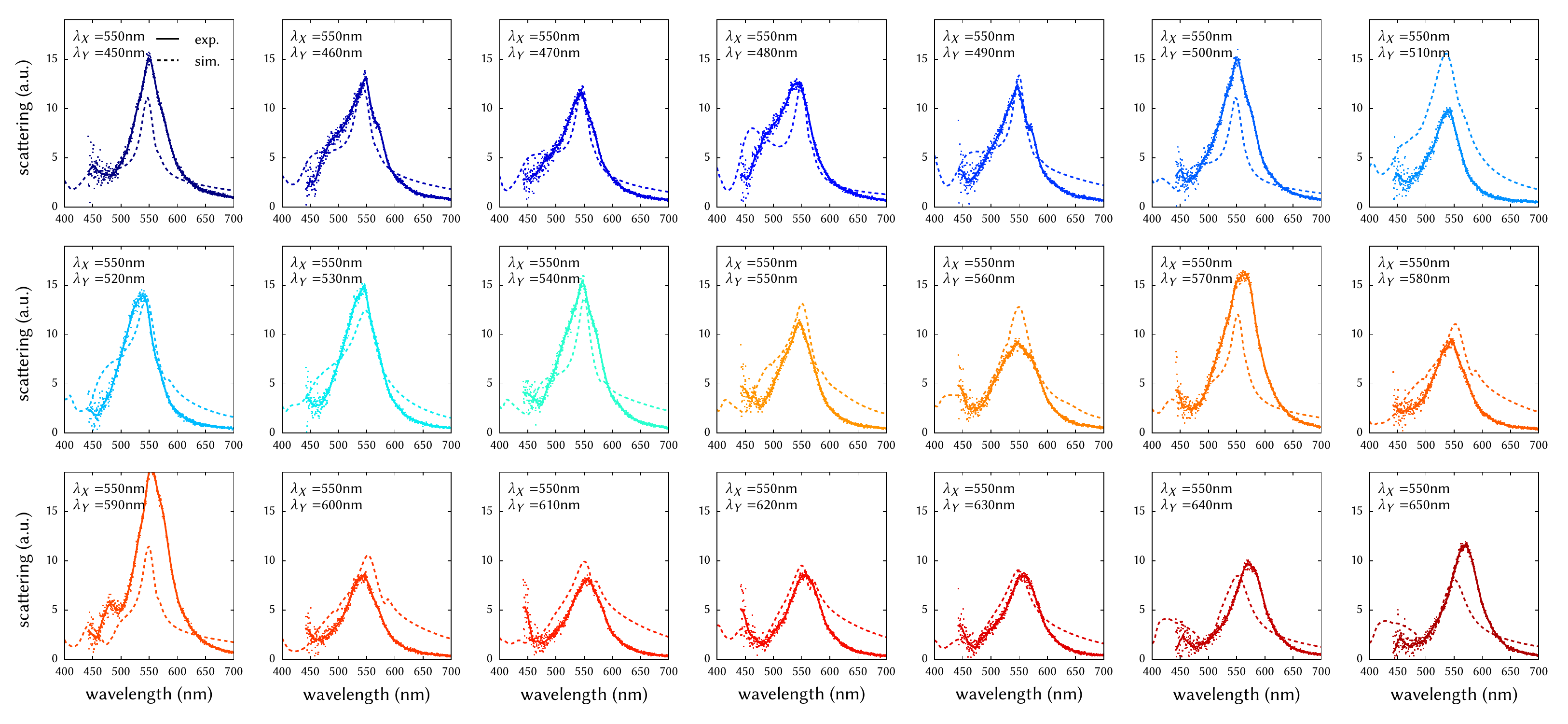} 
\caption{\FIGCAPTIONPREFIX 
\textbf{Comparison simulation / experiment -- \(\mathbf{X}\)-polarisation.} 
Simulated (dashed) and experimental (points/solid line) spectra of EMO-structures for different wavelengths \(\lambda_Y\) at constant \(\lambda_X=550\,\)nm. Polarisation filtered along \(X\). Simulated data: Scattering efficiency, exp. data: Arbitrary units.}
\label{figSI:tuning_comparison_TE}
\end{figure}

\begin{figure}[h!]
\centering
\includegraphics[width=\textwidth, page=1]{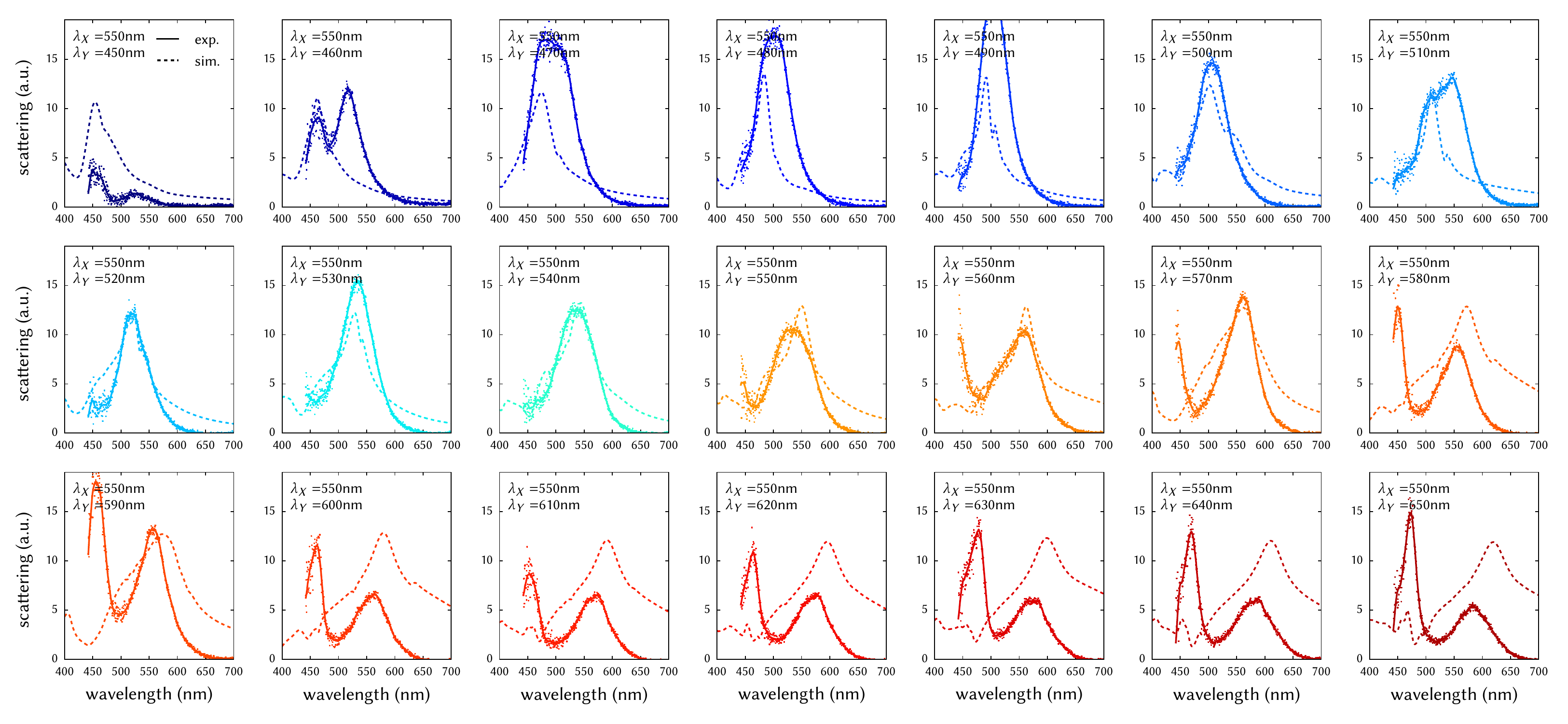} 
\caption{\FIGCAPTIONPREFIX 
\textbf{Comparison simulation / experiment -- \(\mathbf{Y}\)-polarisation.} 
Simulated (dashed) and experimental (points/solid line) spectra of EMO-structures for different wavelengths \(\lambda_Y\) at constant \(\lambda_X=550\,\)nm. Polarisation filtered along \(Y\). Simulated data: Scattering efficiency, exp. data: Arbitrary units.}
\label{figSI:tuning_comparison_TM}
\end{figure}

In order to illustrate the tuning of the optical resonances with the EMO, the experimental spectra in figure~5 of the main paper are normalized to intensity at the red-most maximum.
The quantitative comparison of the measured and calculated spectra, shows a general good agreement. 
Fig.~\ref{figSI:tuning_comparison_TE} shows the case of \(X\)-polarisation with a constant target wavelength of \(\lambda_X=550\,\)nm. 
Fig.~\ref{figSI:tuning_comparison_TM} shows the case of \(Y\)-polarisation (target wavelength tuned from \(\lambda_Y=450\,\)nm to \(\lambda_X=650\,\)nm).
The colours correspond to the coding used in the main paper, figure~5. 

\enlargethispage{1cm}

\clearpage
We attribute the strong enhancement of the scattering around \(\lambda=450\,\)nm, occurring under \(Y\) polarisation for the most red-scattering structures to a cavity effect in the \(145\,\)nm thick SiO\(_2\) layer of the SOI substrate (see also Methods section in main paper).
In the simulations, a bulk SiO\(_2\) substrate is assumed, hence the impact of the finite layer size is not taken into account.

We note that it is in principle possible to take into account multi-layered environments in GDM simulations.\cite{cai_fast_2000, paulus_accurate_2000}
However, the computation of these Green’s Dyads lies out of the scope of the present paper. Furthermore, it would in any case be too computationally demanding and dramatically slow down the optimisation speed. 
Therefore, throughout this work we restricted the calculations to an analytical, non-retarded surface propagator.\cite{girard_shaping_2008}

\begin{figure}[h!]
\centering
\includegraphics[page=1]{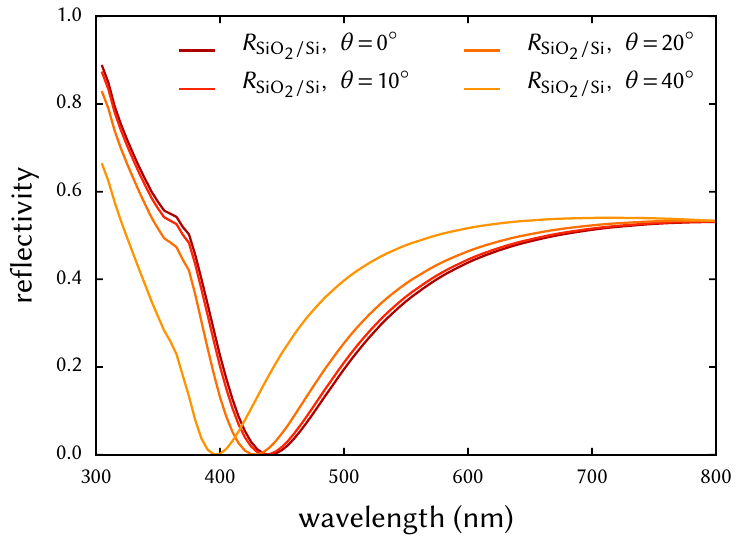} 
\caption{\FIGCAPTIONPREFIX 
\textbf{Reflectivity of Si/SiO\(_2\)/Si Fabry-Perot cavity under different incident angles.} 
Reflectivity of a Si/SiO\(_2\)/Si Fabry-Perot cavity with \(d_{\text{SiO\(_2\)}}=145\,\)nm and \(d_{\text{Si}}=100\,\)nm at the upper upper SiO\(_2\)/Si interface under oblique incidence (polarisation-averaged).}
\label{figSI:reflectivity_fabryperot}
\end{figure}

Figure~\ref{figSI:reflectivity_fabryperot} shows the reflectivity of a Si/SiO\(_2\)/Si layer-stack with thicknesses \(d_{\text{Si-Substrate}}=\infty\), \(d_{\text{SiO\(_2\)}}=145\,\)nm and \(d_{\text{Si}}=100\,\)nm.
While pure SiO\(_2\) has a negligible dispersion in the considered wavelength range (not shown, see \emph{e.g.} Ref.~\onlinecite{malitson_interspecimen_1965}), the reflectivity of the layer-stack is strongly wavelength dependent and a steep minimum around \(\lambda=450\,\)nm occurs. 
For wavelengths larger than \(\lambda \approx 500\,\)nm however, the optical response of the stack is relatively flat and the presence of the underlying cavity does not influence the scattering from the antennas.
To study the impact of oblique incidence (we use a NA\,0.45 objective, corresponding to a maximum incident angle of \(\approx 27^{\circ}\)), we furthermore show the polarisation averaged reflectivity of the cavity for different incident angles.
Up to an angle of \(20^{\circ}\), the reflectivity spectra of the cavity remain almost unchanged and also for higher incident angles the effect is weak.

Owing to the strongly reduced reflectance at the SiO\(_2\)/Si interface around \(\lambda=450\,\)nm, the electric field amplitude is enhanced, resulting in an enhanced scattering of a nano-object placed in this region.
This effect eventually boosts the weak higher-order resonance occurring for the most red-scattering nano-structures (bottom row in Fig.~\ref{figSI:tuning_comparison_TM}), which lies exactly in the spectral region of minimum reflectance at the SiO\(_2\)/Si interface. 
This leads to the observed disagreement between simulation and measurement.

%
%

Due to the above described cavity effect, also in the structures obtained using \(\lambda_X=\lambda_Y=630\,\)nm (main paper figure~3) a relative amplification of the second resonance around \(\lambda=450\,\)nm is occurring in the experimental data when comparing to the simulations.

\clearpage
\subsection{Arrays of structures}

\begin{figure}[h!]
\centering
\includegraphics[width=0.9\textwidth,page=1]{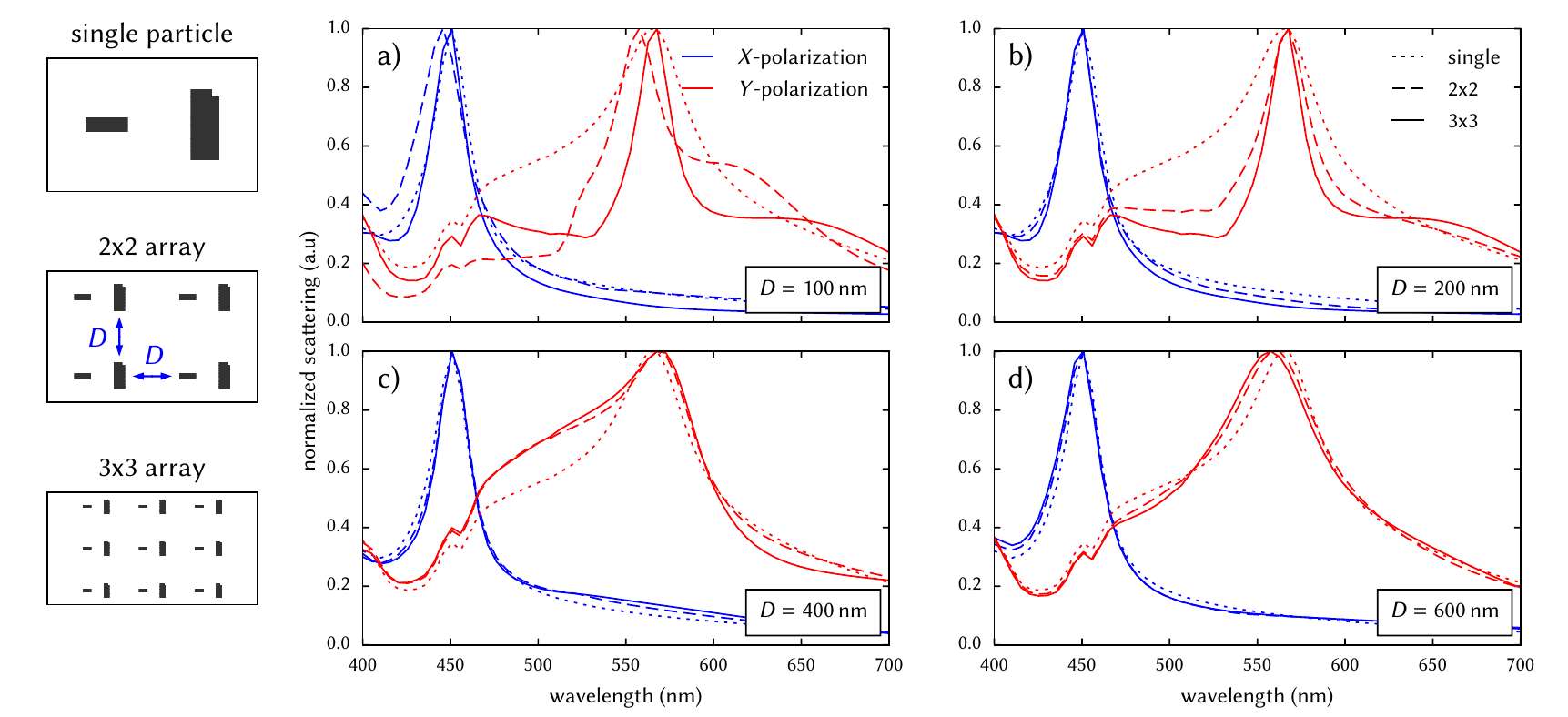} 
\caption{\FIGCAPTIONPREFIX 
\textbf{Impact of array-arrangement of structures on scattering spectra.}
Left: Sketches of the investigated structures. Test-subject is a structure optimised for \(\lambda_X=450\,\)nm and \(\lambda_Y=570\,\)nm. Blue arrows indicate the definition of the antenna spacing \(D\).
Right: Spectra for \(X\)- and \(Y\)-polarised incidence for different values of the spacing. The cases of a single particles, a \(2\times 2\) and a \(3\times 3\) array are compared for
a) \(D=100\,\)nm, b) \(D=200\,\)nm, c) \(D=400\,\)nm and d) \(D=600\,\)nm.}
\label{figSI:particule_array_spectra}
\end{figure}

\begin{figure}[h!]
\centering
\includegraphics[page=1]{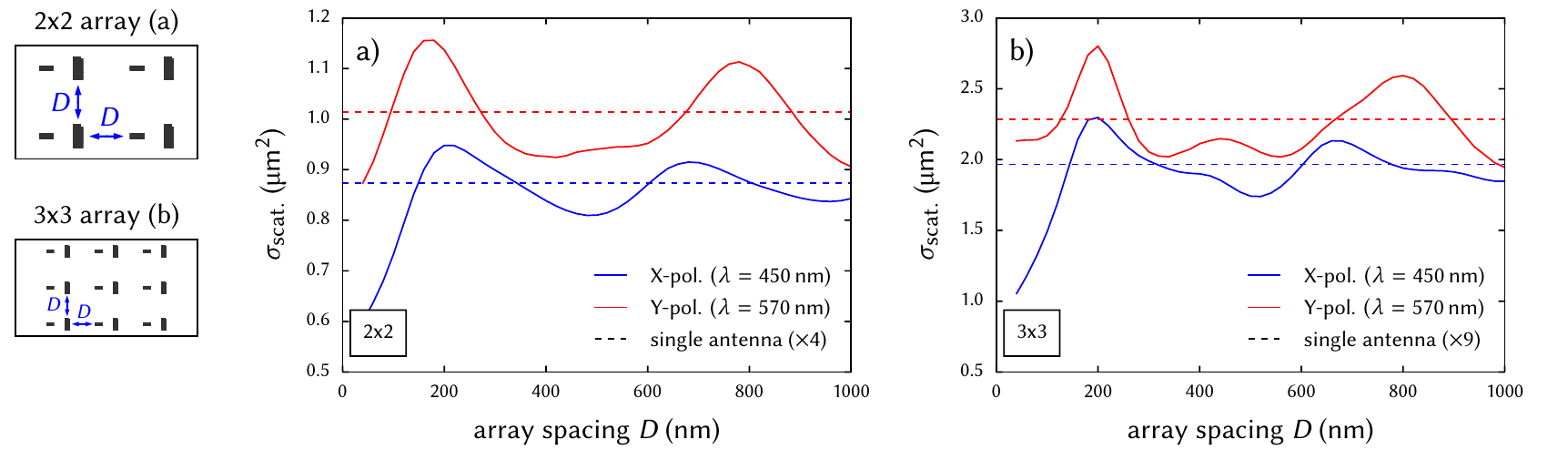} 
\caption{\FIGCAPTIONPREFIX
\textbf{Impact of the density of arrays on their scattering spectra.}
Scattering for \(X\)- and \(Y\)-polarised incidence at the design resonance wavelengths (\(\lambda_X=450\,\)nm and \(\lambda_X=570\,\)nm) as function of the array spacing for a) a \(2\times 2\)-array and b) a \(3\times 3\)-array. 
The scattering of a single structure (\(\times\) the number of repetitions in the array) is plotted as a dashed line.
Left: Illustrations of the investigated arrays (same as in Fig.~\ref{figSI:particule_array_spectra}).
Blue arrows indicate the definition of the spacing~\(D\).}
\label{figSI:particule_array_intensity}
\end{figure}

We want to investigate the effect of arranging the individual particles in sparse arrays (with ``sparse'' we address a very low material-coverage of the surface).
For this purpose, a single antenna is compared to \(2\times 2\) and \(3\times 3\) arrays of the same particle.
Fig.~\ref{figSI:particule_array_spectra} shows spectra for different minimal inter-antenna distances of \(D=100\,\)nm, \(D=200\,\)nm, \(D=400\,\)nm and \(D=600\,\)nm ((a), (b), (c) and (d) respectively).
The overall optical response is only significantly influenced for very small distances \(D\lesssim 400\,\)nm in the considered wavelength range.
The intensities of the peaks are however subject to a far stronger variation, as shown in figure~\ref{figSI:particule_array_intensity}.

In conclusion, to conserve the designed resonance peaks, an interspacing of several \(100\,\)nm should be sufficient (in the mask design, we chose \(D = 250\,\)nm, which we think should in no case be fallen below). 
However, intensity variations must be accounted for at the level of metasurface-design if the scattered intensity is supposed to match a specific target value.

Note, that it is possible to take the arrangement of nano-structures in (finite-size) arrays into account at the GDM computation level using an appropriate Green's Dyad.\cite{chaumet_simulation_2009, gallinet_electromagnetic_2009}
\enlargethispage{1cm}

%
%
%

\begin{thebibliography}{10}
\expandafter\ifx\csname url\endcsname\relax
  \def\url#1{\texttt{#1}}\fi
\expandafter\ifx\csname urlprefix\endcsname\relax\def\urlprefix{URL }\fi
\providecommand{\bibinfo}[2]{#2}
\providecommand{\eprint}[2][]{\url{#2}}

\bibitem{tan_plasmonic_2014}
\bibinfo{author}{Tan, S.~J.} \emph{et~al.}
\newblock \bibinfo{title}{Plasmonic {Color} {Palettes} for {Photorealistic}
  {Printing} with {Aluminum} {Nanostructures}}.
\newblock \emph{\bibinfo{journal}{Nano Lett.}} \textbf{\bibinfo{volume}{14}},
  \bibinfo{pages}{4023--4029} (\bibinfo{year}{2014}).
\newblock \urlprefix\url{http://dx.doi.org/10.1021/nl501460x}.

\bibitem{lindfors_imaging_2016}
\bibinfo{author}{Lindfors, K.} \emph{et~al.}
\newblock \bibinfo{title}{Imaging and {Steering} {Unidirectional} {Emission}
  from {Nanoantenna} {Array} {Metasurfaces}}.
\newblock \emph{\bibinfo{journal}{ACS Photonics}} \textbf{\bibinfo{volume}{3}},
  \bibinfo{pages}{286--292} (\bibinfo{year}{2016}).
\newblock \urlprefix\url{http://dx.doi.org/10.1021/acsphotonics.5b00646}.

\bibitem{black_optimal_2014}
\bibinfo{author}{Black, L.-J.}, \bibinfo{author}{Wang, Y.},
  \bibinfo{author}{de~Groot, C.~H.}, \bibinfo{author}{Arbouet, A.} \&
  \bibinfo{author}{Muskens, O.~L.}
\newblock \bibinfo{title}{Optimal {Polarization} {Conversion} in {Coupled}
  {Dimer} {Plasmonic} {Nanoantennas} for {Metasurfaces}}.
\newblock \emph{\bibinfo{journal}{ACS Nano}} \textbf{\bibinfo{volume}{8}},
  \bibinfo{pages}{6390--6399} (\bibinfo{year}{2014}).
\newblock \urlprefix\url{http://dx.doi.org/10.1021/nn501889s}.

\bibitem{valev_chirality_2013}
\bibinfo{author}{Valev, V.~K.}, \bibinfo{author}{Baumberg, J.~J.},
  \bibinfo{author}{Sibilia, C.} \& \bibinfo{author}{Verbiest, T.}
\newblock \bibinfo{title}{Chirality and {Chiroptical} {Effects} in {Plasmonic}
  {Nanostructures}: {Fundamentals}, {Recent} {Progress}, and {Outlook}}.
\newblock \emph{\bibinfo{journal}{Adv. Mater.}} \textbf{\bibinfo{volume}{25}},
  \bibinfo{pages}{2517--2534} (\bibinfo{year}{2013}).
\newblock
  \urlprefix\url{http://onlinelibrary.wiley.com/doi/10.1002/adma.201205178/abstract}.

\bibitem{kauranen_nonlinear_2012}
\bibinfo{author}{Kauranen, M.} \& \bibinfo{author}{Zayats, A.~V.}
\newblock \bibinfo{title}{Nonlinear plasmonics}.
\newblock \emph{\bibinfo{journal}{Nat Photon}} \textbf{\bibinfo{volume}{6}},
  \bibinfo{pages}{737--748} (\bibinfo{year}{2012}).
\newblock
  \urlprefix\url{http://www.nature.com/nphoton/journal/v6/n11/full/nphoton.2012.244.html}.

\bibitem{albella_electric_2014}
\bibinfo{author}{Albella, P.}, \bibinfo{author}{Alcaraz de~la Osa, R.},
  \bibinfo{author}{Moreno, F.} \& \bibinfo{author}{Maier, S.~A.}
\newblock \bibinfo{title}{Electric and {Magnetic} {Field} {Enhancement} with
  {Ultralow} {Heat} {Radiation} {Dielectric} {Nanoantennas}: {Considerations}
  for {Surface}-{Enhanced} {Spectroscopies}}.
\newblock \emph{\bibinfo{journal}{ACS Photonics}} \textbf{\bibinfo{volume}{1}},
  \bibinfo{pages}{524--529} (\bibinfo{year}{2014}).
\newblock \urlprefix\url{http://dx.doi.org/10.1021/ph500060s}.

\bibitem{bakker_magnetic_2015}
\bibinfo{author}{Bakker, R.~M.} \emph{et~al.}
\newblock \bibinfo{title}{Magnetic and {Electric} {Hotspots} with {Silicon}
  {Nanodimers}}.
\newblock \emph{\bibinfo{journal}{Nano Lett.}} \textbf{\bibinfo{volume}{15}},
  \bibinfo{pages}{2137--2142} (\bibinfo{year}{2015}).
\newblock \urlprefix\url{http://dx.doi.org/10.1021/acs.nanolett.5b00128}.

\bibitem{ginn_realizing_2012}
\bibinfo{author}{Ginn, J.~C.} \emph{et~al.}
\newblock \bibinfo{title}{Realizing {Optical} {Magnetism} from {Dielectric}
  {Metamaterials}}.
\newblock \emph{\bibinfo{journal}{Phys. Rev. Lett.}}
  \textbf{\bibinfo{volume}{108}}, \bibinfo{pages}{097402}
  (\bibinfo{year}{2012}).
\newblock
  \urlprefix\url{http://link.aps.org/doi/10.1103/PhysRevLett.108.097402}.

\bibitem{kuznetsov_magnetic_2012}
\bibinfo{author}{Kuznetsov, A.~I.}, \bibinfo{author}{Miroshnichenko, A.~E.},
  \bibinfo{author}{Fu, Y.~H.}, \bibinfo{author}{Zhang, J.} \&
  \bibinfo{author}{Luk’yanchuk, B.}
\newblock \bibinfo{title}{Magnetic light}.
\newblock \emph{\bibinfo{journal}{Scientific Reports}}
  \textbf{\bibinfo{volume}{2}}, \bibinfo{pages}{492} (\bibinfo{year}{2012}).
\newblock \urlprefix\url{http://www.nature.com/articles/srep00492}.

\bibitem{schmidt_dielectric_2012}
\bibinfo{author}{Schmidt, M.~K.} \emph{et~al.}
\newblock \bibinfo{title}{Dielectric antennas - a suitable platform for
  controlling magnetic dipolar emission}.
\newblock \emph{\bibinfo{journal}{Optics Express}}
  \textbf{\bibinfo{volume}{20}}, \bibinfo{pages}{13636} (\bibinfo{year}{2012}).
\newblock
  \urlprefix\url{https://www.osapublishing.org/oe/abstract.cfm?uri=oe-20-13-13636}.

\bibitem{cao_tuning_2010}
\bibinfo{author}{Cao, L.}, \bibinfo{author}{Fan, P.}, \bibinfo{author}{Barnard,
  E.~S.}, \bibinfo{author}{Brown, A.~M.} \& \bibinfo{author}{Brongersma, M.~L.}
\newblock \bibinfo{title}{Tuning the {Color} of {Silicon} {Nanostructures}}.
\newblock \emph{\bibinfo{journal}{Nano Lett.}} \textbf{\bibinfo{volume}{10}},
  \bibinfo{pages}{2649--2654} (\bibinfo{year}{2010}).
\newblock \urlprefix\url{http://dx.doi.org/10.1021/nl1013794}.

\bibitem{traviss_antenna_2015}
\bibinfo{author}{Traviss, D.~J.}, \bibinfo{author}{Schmidt, M.~K.},
  \bibinfo{author}{Aizpurua, J.} \& \bibinfo{author}{Muskens, O.~L.}
\newblock \bibinfo{title}{Antenna resonances in low aspect ratio semiconductor
  nanowires}.
\newblock \emph{\bibinfo{journal}{Optics Express}}
  \textbf{\bibinfo{volume}{23}}, \bibinfo{pages}{22771} (\bibinfo{year}{2015}).
\newblock
  \urlprefix\url{https://www.osapublishing.org/abstract.cfm?URI=oe-23-17-22771}.

\bibitem{zhao_full-color_2016}
\bibinfo{author}{Zhao, W.} \emph{et~al.}
\newblock \bibinfo{title}{Full-color hologram using spatial multiplexing of
  dielectric metasurface}.
\newblock \emph{\bibinfo{journal}{Optics Letters}}
  \textbf{\bibinfo{volume}{41}}, \bibinfo{pages}{147} (\bibinfo{year}{2016}).
\newblock
  \urlprefix\url{https://www.osapublishing.org/abstract.cfm?URI=ol-41-1-147}.

\bibitem{yu_high-transmission_2015}
\bibinfo{author}{Yu, Y.~F.} \emph{et~al.}
\newblock \bibinfo{title}{High-transmission dielectric metasurface with 2pi
  phase control at visible wavelengths}.
\newblock \emph{\bibinfo{journal}{Laser \& Photonics Reviews}}
  \textbf{\bibinfo{volume}{9}}, \bibinfo{pages}{412--418}
  (\bibinfo{year}{2015}).
\newblock
  \urlprefix\url{http://onlinelibrary.wiley.com/doi/10.1002/lpor.201500041/abstract}.

\bibitem{shcherbakov_enhanced_2014}
\bibinfo{author}{Shcherbakov, M.~R.} \emph{et~al.}
\newblock \bibinfo{title}{Enhanced {Third}-{Harmonic} {Generation} in {Silicon}
  {Nanoparticles} {Driven} by {Magnetic} {Response}}.
\newblock \emph{\bibinfo{journal}{Nano Lett.}} \textbf{\bibinfo{volume}{14}},
  \bibinfo{pages}{6488--6492} (\bibinfo{year}{2014}).
\newblock \urlprefix\url{http://dx.doi.org/10.1021/nl503029j}.

\bibitem{wiecha_enhanced_2015}
\bibinfo{author}{Wiecha, P.~R.} \emph{et~al.}
\newblock \bibinfo{title}{Enhanced nonlinear optical response from individual
  silicon nanowires}.
\newblock \emph{\bibinfo{journal}{Phys. Rev. B}} \textbf{\bibinfo{volume}{91}},
  \bibinfo{pages}{121416} (\bibinfo{year}{2015}).
\newblock \urlprefix\url{http://link.aps.org/doi/10.1103/PhysRevB.91.121416}.

\bibitem{fu_directional_2013}
\bibinfo{author}{Fu, Y.~H.}, \bibinfo{author}{Kuznetsov, A.~I.},
  \bibinfo{author}{Miroshnichenko, A.~E.}, \bibinfo{author}{Yu, Y.~F.} \&
  \bibinfo{author}{Luk’yanchuk, B.}
\newblock \bibinfo{title}{Directional visible light scattering by silicon
  nanoparticles}.
\newblock \emph{\bibinfo{journal}{Nat Commun}} \textbf{\bibinfo{volume}{4}},
  \bibinfo{pages}{1527} (\bibinfo{year}{2013}).
\newblock
  \urlprefix\url{http://www.nature.com/ncomms/journal/v4/n2/full/ncomms2538.html}.

\bibitem{sivanandam_introduction_2008}
\bibinfo{author}{Sivanandam, S.} \& \bibinfo{author}{Deepa, S.}
\newblock \emph{\bibinfo{title}{Introduction to {Genetic} {Algorithms}}}
  (\bibinfo{publisher}{Springer}, \bibinfo{address}{Heidelberg},
  \bibinfo{year}{2008}).
\newblock
  \urlprefix\url{http://link.springer.com/book/10.1007%2F978-3-540-73190-0}.

\bibitem{forestiere_particle-swarm_2010}
\bibinfo{author}{Forestiere, C.} \emph{et~al.}
\newblock \bibinfo{title}{Particle-swarm optimization of broadband
  nanoplasmonic arrays}.
\newblock \emph{\bibinfo{journal}{Optics Letters}}
  \textbf{\bibinfo{volume}{35}}, \bibinfo{pages}{133} (\bibinfo{year}{2010}).
\newblock
  \urlprefix\url{https://www.osapublishing.org/ol/abstract.cfm?uri=ol-35-2-133}.

\bibitem{feichtner_evolutionary_2012}
\bibinfo{author}{Feichtner, T.}, \bibinfo{author}{Selig, O.},
  \bibinfo{author}{Kiunke, M.} \& \bibinfo{author}{Hecht, B.}
\newblock \bibinfo{title}{Evolutionary {Optimization} of {Optical} {Antennas}}.
\newblock \emph{\bibinfo{journal}{Phys. Rev. Lett.}}
  \textbf{\bibinfo{volume}{109}}, \bibinfo{pages}{127701}
  (\bibinfo{year}{2012}).
\newblock
  \urlprefix\url{http://link.aps.org/doi/10.1103/PhysRevLett.109.127701}.

\bibitem{forestiere_genetically_2012}
\bibinfo{author}{Forestiere, C.} \emph{et~al.}
\newblock \bibinfo{title}{Genetically {Engineered} {Plasmonic} {Nanoarrays}}.
\newblock \emph{\bibinfo{journal}{Nano Lett.}} \textbf{\bibinfo{volume}{12}},
  \bibinfo{pages}{2037--2044} (\bibinfo{year}{2012}).
\newblock \urlprefix\url{http://dx.doi.org/10.1021/nl300140g}.

\bibitem{forestiere_inverse_2016}
\bibinfo{author}{Forestiere, C.}, \bibinfo{author}{He, Y.},
  \bibinfo{author}{Wang, R.}, \bibinfo{author}{Kirby, R.~M.} \&
  \bibinfo{author}{Dal~Negro, L.}
\newblock \bibinfo{title}{Inverse {Design} of {Metal} {Nanoparticles}’
  {Morphology}}.
\newblock \emph{\bibinfo{journal}{ACS Photonics}} \textbf{\bibinfo{volume}{3}},
  \bibinfo{pages}{68--78} (\bibinfo{year}{2016}).
\newblock \urlprefix\url{http://dx.doi.org/10.1021/acsphotonics.5b00463}.

\bibitem{ginzburg_resonances_2011}
\bibinfo{author}{Ginzburg, P.}, \bibinfo{author}{Berkovitch, N.},
  \bibinfo{author}{Nevet, A.}, \bibinfo{author}{Shor, I.} \&
  \bibinfo{author}{Orenstein, M.}
\newblock \bibinfo{title}{Resonances {On}-{Demand} for {Plasmonic}
  {Nano}-{Particles}}.
\newblock \emph{\bibinfo{journal}{Nano Lett.}} \textbf{\bibinfo{volume}{11}},
  \bibinfo{pages}{2329--2333} (\bibinfo{year}{2011}).
\newblock \urlprefix\url{http://dx.doi.org/10.1021/nl200612f}.

\bibitem{macias_heuristic_2012}
\bibinfo{author}{Macías, D.}, \bibinfo{author}{Adam, P.-M.},
  \bibinfo{author}{Ruíz-Cortés, V.}, \bibinfo{author}{Rodríguez-Oliveros,
  R.} \& \bibinfo{author}{Sánchez-Gil, J.~A.}
\newblock \bibinfo{title}{Heuristic optimization for the design of plasmonic
  nanowires with specific resonant and scattering properties}.
\newblock \emph{\bibinfo{journal}{Optics Express}}
  \textbf{\bibinfo{volume}{20}}, \bibinfo{pages}{13146} (\bibinfo{year}{2012}).
\newblock
  \urlprefix\url{https://www.osapublishing.org/oe/abstract.cfm?uri=oe-20-12-13146}.

\bibitem{bigourdan_design_2014}
\bibinfo{author}{Bigourdan, F.}, \bibinfo{author}{Marquier, F.},
  \bibinfo{author}{Hugonin, J.-P.} \& \bibinfo{author}{Greffet, J.-J.}
\newblock \bibinfo{title}{Design of highly efficient metallo-dielectric patch
  antennas for single-photon emission}.
\newblock \emph{\bibinfo{journal}{Optics Express}}
  \textbf{\bibinfo{volume}{22}}, \bibinfo{pages}{2337} (\bibinfo{year}{2014}).
\newblock
  \urlprefix\url{https://www.osapublishing.org/oe/abstract.cfm?uri=oe-22-3-2337}.

\bibitem{mirzaei_superscattering_2014}
\bibinfo{author}{Mirzaei, A.}, \bibinfo{author}{Miroshnichenko, A.~E.},
  \bibinfo{author}{Shadrivov, I.~V.} \& \bibinfo{author}{Kivshar, Y.~S.}
\newblock \bibinfo{title}{Superscattering of light optimized by a genetic
  algorithm}.
\newblock \emph{\bibinfo{journal}{Applied Physics Letters}}
  \textbf{\bibinfo{volume}{105}}, \bibinfo{pages}{011109}
  (\bibinfo{year}{2014}).
\newblock
  \urlprefix\url{http://scitation.aip.org/content/aip/journal/apl/105/1/10.1063/1.4887475}.

\bibitem{chen_optimal_2007}
\bibinfo{author}{Chen, P.~Y.}, \bibinfo{author}{Chen, C.~H.},
  \bibinfo{author}{Wu, J.~S.}, \bibinfo{author}{Wen, H.~C.} \&
  \bibinfo{author}{Wang, W.~P.}
\newblock \bibinfo{title}{Optimal design of integrally gated {CNT}
  field-emission devices using a genetic algorithm}.
\newblock \emph{\bibinfo{journal}{Nanotechnology}}
  \textbf{\bibinfo{volume}{18}}, \bibinfo{pages}{395203}
  (\bibinfo{year}{2007}).
\newblock
  \urlprefix\url{http://stacks.iop.org/0957-4484/18/i=39/a=395203?key=crossref.d374430889ccb7bbffbf78b97103aeb0}.

\bibitem{mengyu_wang_optimization_2014}
\bibinfo{author}{{Mengyu Wang}}, \bibinfo{author}{{Aytac Alparslan}},
  \bibinfo{author}{{Sascha M. Schnepp}} \& \bibinfo{author}{{Christian
  Hafner}}.
\newblock \bibinfo{title}{Optimization of a {Plasmon}-{Assisted} {Waveguide}
  {Coupler} {Using} {FEM} and {MMP}}.
\newblock \emph{\bibinfo{journal}{Progress In Electromagnetics Research B}}
  \textbf{\bibinfo{volume}{59}}, \bibinfo{pages}{219--229}
  (\bibinfo{year}{2014}).

\bibitem{kessentini_particle_2011}
\bibinfo{author}{Kessentini, S.}, \bibinfo{author}{Barchiesi, D.},
  \bibinfo{author}{Grosges, T.} \& \bibinfo{author}{Lamy de~la Chapelle, M.}
\newblock \bibinfo{title}{Particle swarm optimization and evolutionary methods
  for plasmonic biomedical applications}.
\newblock In \emph{\bibinfo{booktitle}{2011 {IEEE} {Congress} on {Evolutionary}
  {Computation} ({CEC})}}, \bibinfo{pages}{2315--2320} (\bibinfo{year}{2011}).

\bibitem{jung_robust_2016}
\bibinfo{author}{Jung, J.}
\newblock \bibinfo{title}{Robust {Design} of {Plasmonic} {Waveguide} {Using}
  {Gradient} {Index} and {Multiobjective} {Optimization}}.
\newblock \emph{\bibinfo{journal}{IEEE Photonics Technology Letters}}
  \textbf{\bibinfo{volume}{28}}, \bibinfo{pages}{756--758}
  (\bibinfo{year}{2016}).

\bibitem{deb_multi-objective_2001}
\bibinfo{author}{Deb, K.}
\newblock \emph{\bibinfo{title}{Multi-objective optimization using evolutionary
  algorithms}}, vol.~\bibinfo{volume}{16} (\bibinfo{publisher}{Wiley},
  \bibinfo{year}{2001}).

\bibitem{shegai_bimetallic_2011}
\bibinfo{author}{Shegai, T.} \emph{et~al.}
\newblock \bibinfo{title}{A bimetallic nanoantenna for directional colour
  routing}.
\newblock \emph{\bibinfo{journal}{Nat Commun}} \textbf{\bibinfo{volume}{2}},
  \bibinfo{pages}{481} (\bibinfo{year}{2011}).
\newblock
  \urlprefix\url{http://www.nature.com/ncomms/journal/v2/n9/full/ncomms1490.html}.

\bibitem{aouani_ultrasensitive_2013}
\bibinfo{author}{Aouani, H.} \emph{et~al.}
\newblock \bibinfo{title}{Ultrasensitive {Broadband} {Probing} of {Molecular}
  {Vibrational} {Modes} with {Multifrequency} {Optical} {Antennas}}.
\newblock \emph{\bibinfo{journal}{ACS Nano}} \textbf{\bibinfo{volume}{7}},
  \bibinfo{pages}{669--675} (\bibinfo{year}{2013}).
\newblock \urlprefix\url{http://dx.doi.org/10.1021/nn304860t}.

\bibitem{dopf_coupled_2015}
\bibinfo{author}{Dopf, K.} \emph{et~al.}
\newblock \bibinfo{title}{Coupled {T}-{Shaped} {Optical} {Antennas} with {Two}
  {Resonances} {Localized} in a {Common} {Nanogap}}.
\newblock \emph{\bibinfo{journal}{ACS Photonics}} \textbf{\bibinfo{volume}{2}},
  \bibinfo{pages}{1644--1651} (\bibinfo{year}{2015}).
\newblock \urlprefix\url{http://dx.doi.org/10.1021/acsphotonics.5b00446}.

\bibitem{giannini_excitation_2008}
\bibinfo{author}{Giannini, V.} \& \bibinfo{author}{Sánchez-Gil, J.~A.}
\newblock \bibinfo{title}{Excitation and emission enhancement of single
  molecule fluorescence through multiple surface-plasmon resonances on metal
  trimer nanoantennas}.
\newblock \emph{\bibinfo{journal}{Optics Letters}}
  \textbf{\bibinfo{volume}{33}}, \bibinfo{pages}{899} (\bibinfo{year}{2008}).
\newblock
  \urlprefix\url{https://www.osapublishing.org/abstract.cfm?URI=ol-33-9-899}.

\bibitem{harutyunyan_enhancing_2012}
\bibinfo{author}{Harutyunyan, H.}, \bibinfo{author}{Volpe, G.},
  \bibinfo{author}{Quidant, R.} \& \bibinfo{author}{Novotny, L.}
\newblock \bibinfo{title}{Enhancing the {Nonlinear} {Optical} {Response}
  {Using} {Multifrequency} {Gold}-{Nanowire} {Antennas}}.
\newblock \emph{\bibinfo{journal}{Phys. Rev. Lett.}}
  \textbf{\bibinfo{volume}{108}}, \bibinfo{pages}{217403}
  (\bibinfo{year}{2012}).
\newblock
  \urlprefix\url{http://link.aps.org/doi/10.1103/PhysRevLett.108.217403}.

\bibitem{celebrano_mode_2015}
\bibinfo{author}{Celebrano, M.} \emph{et~al.}
\newblock \bibinfo{title}{Mode matching in multiresonant plasmonic nanoantennas
  for enhanced second harmonic generation}.
\newblock \emph{\bibinfo{journal}{Nat Nano}} \textbf{\bibinfo{volume}{10}},
  \bibinfo{pages}{412--417} (\bibinfo{year}{2015}).
\newblock
  \urlprefix\url{http://www.nature.com/nnano/journal/v10/n5/full/nnano.2015.69.html}.

\bibitem{girard_near_2005}
\bibinfo{author}{Girard, C.}
\newblock \bibinfo{title}{Near fields in nanostructures}.
\newblock \emph{\bibinfo{journal}{Reports on Progress in Physics}}
  \textbf{\bibinfo{volume}{68}}, \bibinfo{pages}{1883--1933}
  (\bibinfo{year}{2005}).
\newblock
  \urlprefix\url{http://stacks.iop.org/0034-4885/68/i=8/a=R05?key=crossref.0a49722c35b5f8f658773d36f47aa4be}.

\bibitem{goh_comparative_2016}
\bibinfo{author}{Goh, X.~M.}, \bibinfo{author}{Ng, R. J.~H.},
  \bibinfo{author}{Wang, S.}, \bibinfo{author}{Tan, S.~J.} \&
  \bibinfo{author}{Yang, J.~K.}
\newblock \bibinfo{title}{Comparative {Study} of {Plasmonic} {Colors} from
  {All}-{Metal} {Structures} of {Posts} and {Pits}}.
\newblock \emph{\bibinfo{journal}{ACS Photonics}} \textbf{\bibinfo{volume}{3}},
  \bibinfo{pages}{1000--1009} (\bibinfo{year}{2016}).
\newblock \urlprefix\url{http://dx.doi.org/10.1021/acsphotonics.6b00099}.

\bibitem{li_dual_2016}
\bibinfo{author}{Li, Z.}, \bibinfo{author}{Clark, A.~W.} \&
  \bibinfo{author}{Cooper, J.~M.}
\newblock \bibinfo{title}{Dual {Color} {Plasmonic} {Pixels} {Create} a
  {Polarization} {Controlled} {Nano} {Color} {Palette}}.
\newblock \emph{\bibinfo{journal}{ACS Nano}} \textbf{\bibinfo{volume}{10}},
  \bibinfo{pages}{492--498} (\bibinfo{year}{2016}).
\newblock \urlprefix\url{http://dx.doi.org/10.1021/acsnano.5b05411}.

\bibitem{beume_sms-emoa:_2007}
\bibinfo{author}{Beume, N.}, \bibinfo{author}{Naujoks, B.} \&
  \bibinfo{author}{Emmerich, M.}
\newblock \bibinfo{title}{{SMS}-{EMOA}: {Multiobjective} selection based on
  dominated hypervolume}.
\newblock \emph{\bibinfo{journal}{European Journal of Operational Research}}
  \textbf{\bibinfo{volume}{181}}, \bibinfo{pages}{1653--1669}
  (\bibinfo{year}{2007}).
\newblock
  \urlprefix\url{http://www.sciencedirect.com/science/article/pii/S0377221706005443}.

\bibitem{martin_generalized_1995}
\bibinfo{author}{Martin, O. J.~F.}, \bibinfo{author}{Girard, C.} \&
  \bibinfo{author}{Dereux, A.}
\newblock \bibinfo{title}{Generalized {Field} {Propagator} for
  {Electromagnetic} {Scattering} and {Light} {Confinement}}.
\newblock \emph{\bibinfo{journal}{Phys. Rev. Lett.}}
  \textbf{\bibinfo{volume}{74}}, \bibinfo{pages}{526--529}
  (\bibinfo{year}{1995}).
\newblock \urlprefix\url{http://link.aps.org/doi/10.1103/PhysRevLett.74.526}.

\bibitem{chaumet_simulation_2009}
\bibinfo{author}{Chaumet, P.~C.} \& \bibinfo{author}{Sentenac, A.}
\newblock \bibinfo{title}{Simulation of light scattering by multilayer
  cross-gratings with the coupled dipole method}.
\newblock \emph{\bibinfo{journal}{Journal of Quantitative Spectroscopy and
  Radiative Transfer}} \textbf{\bibinfo{volume}{110}},
  \bibinfo{pages}{409--414} (\bibinfo{year}{2009}).
\newblock
  \urlprefix\url{http://www.sciencedirect.com/science/article/pii/S0022407308002707}.

\bibitem{gallinet_electromagnetic_2009}
\bibinfo{author}{Gallinet, B.} \& \bibinfo{author}{Martin, O. J.~F.}
\newblock \bibinfo{title}{Electromagnetic {Scattering} of {Finite} and
  {Infinite} 3d {Lattices} in {Polarizable} {Backgrounds}}.
\newblock \emph{\bibinfo{journal}{Theoretical And Computational Nanophotonics
  (Tacona-Photonics 2009)}} \textbf{\bibinfo{volume}{1176}},
  \bibinfo{pages}{63--65} (\bibinfo{year}{2009}).
\newblock \urlprefix\url{http://infoscience.epfl.ch/record/164792}.

\bibitem{biscani_global_2010}
\bibinfo{author}{Biscani, F.}, \bibinfo{author}{Izzo, D.} \&
  \bibinfo{author}{Yam, C.~H.}
\newblock \bibinfo{title}{A {Global} {Optimisation} {Toolbox} for {Massively}
  {Parallel} {Engineering} {Optimisation}}.
\newblock \emph{\bibinfo{journal}{arXiv:1004.3824 [cs, math]}}
  (\bibinfo{year}{2010}).
\newblock \urlprefix\url{http://arxiv.org/abs/1004.3824}.
\newblock \bibinfo{note}{ArXiv: 1004.3824}.

\bibitem{palik_silicon_1997}
\bibinfo{author}{Edwards, D.~F.}
\newblock \bibinfo{title}{Silicon ({Si})*}.
\newblock In \bibinfo{editor}{Palik, E.~D.} (ed.)
  \emph{\bibinfo{booktitle}{Handbook of {Optical} {Constants} of {Solids}}},
  \bibinfo{pages}{547 -- 569} (\bibinfo{publisher}{Academic Press},
  \bibinfo{address}{Burlington}, \bibinfo{year}{1997}).
\newblock
  \urlprefix\url{http://www.sciencedirect.com/science/article/pii/B9780125444156500273}.

\bibitem{girard_shaping_2008}
\bibinfo{author}{Girard, C.}, \bibinfo{author}{Dujardin, E.},
  \bibinfo{author}{Baffou, G.} \& \bibinfo{author}{Quidant, R.}
\newblock \bibinfo{title}{Shaping and manipulation of light fields with
  bottom-up plasmonic structures}.
\newblock \emph{\bibinfo{journal}{New J. Phys.}} \textbf{\bibinfo{volume}{10}},
  \bibinfo{pages}{105016} (\bibinfo{year}{2008}).
\newblock \urlprefix\url{http://iopscience.iop.org/1367-2630/10/10/105016}.

\bibitem{draine_discrete-dipole_1988}
\bibinfo{author}{Draine, B.~T.}
\newblock \bibinfo{title}{The {Discrete}-{Dipole} {Approximation} and its
  {Application} to {Interstellar} {Graphite} {Grains}}.
\newblock \emph{\bibinfo{journal}{Astrophys. J.}}
  \textbf{\bibinfo{volume}{333}}, \bibinfo{pages}{848--872}
  (\bibinfo{year}{1988}).

\bibitem{han_realization_2011}
\bibinfo{author}{Han, X.-L.}, \bibinfo{author}{Larrieu, G.},
  \bibinfo{author}{Fazzini, P.-F.} \& \bibinfo{author}{Dubois, E.}
\newblock \bibinfo{title}{Realization of ultra dense arrays of vertical silicon
  nanowires with defect free surface and perfect anisotropy using a top-down
  approach}.
\newblock \emph{\bibinfo{journal}{Microelectronic Engineering}}
  \textbf{\bibinfo{volume}{88}}, \bibinfo{pages}{2622--2624}
  (\bibinfo{year}{2011}).
\newblock
  \urlprefix\url{http://www.sciencedirect.com/science/article/pii/S0167931710005903}.

\bibitem{guerfi_high_2013}
\bibinfo{author}{Guerfi, Y.}, \bibinfo{author}{Carcenac, F.} \&
  \bibinfo{author}{Larrieu, G.}
\newblock \bibinfo{title}{High resolution {HSQ} nanopillar arrays with low
  energy electron beam lithography}.
\newblock \emph{\bibinfo{journal}{Microelectronic Engineering}}
  \textbf{\bibinfo{volume}{110}}, \bibinfo{pages}{173--176}
  (\bibinfo{year}{2013}).
\newblock
  \urlprefix\url{http://www.sciencedirect.com/science/article/pii/S0167931713002724}.

\end{thebibliography}

\begin{thebibliography}{10}
\expandafter\ifx\csname url\endcsname\relax
  \def\url#1{\texttt{#1}}\fi
\expandafter\ifx\csname urlprefix\endcsname\relax\def\urlprefix{URL }\fi
\providecommand{\bibinfo}[2]{#2}
\providecommand{\eprint}[2][]{\url{#2}}

\bibitem{fliege_steepest_2000}
\bibinfo{author}{Fliege, J.} \& \bibinfo{author}{Svaiter, B.~F.}
\newblock \bibinfo{title}{Steepest descent methods for multicriteria
  optimization}.
\newblock \emph{\bibinfo{journal}{Mathematical Methods of OR}}
  \textbf{\bibinfo{volume}{51}}, \bibinfo{pages}{479--494}
  (\bibinfo{year}{2000}).
\newblock
  \urlprefix\url{http://link.springer.com/article/10.1007/s001860000043}.

\bibitem{giacomini_comparison_2014}
\bibinfo{author}{Giacomini, M.}, \bibinfo{author}{Désidéri, J.-A.} \&
  \bibinfo{author}{Duvigneau, R.}
\newblock \bibinfo{title}{Comparison of multiobjective gradient-based methods
  for structural shape optimization}.
\newblock \bibinfo{type}{report}, \bibinfo{institution}{INRIA}
  (\bibinfo{year}{2014}).
\newblock \urlprefix\url{https://hal.inria.fr/hal-00967601/document}.

\bibitem{desideri_multiple-gradient_2014}
\bibinfo{author}{Désidéri, J.-A.}
\newblock \bibinfo{title}{Multiple-gradient {Descent} {Algorithm} for
  {Pareto}-{Front} {Identification}}.
\newblock In \bibinfo{editor}{Fitzgibbon, W.}, \bibinfo{editor}{Kuznetsov,
  Y.~A.}, \bibinfo{editor}{Neittaanmäki, P.} \& \bibinfo{editor}{Pironneau,
  O.} (eds.) \emph{\bibinfo{booktitle}{Modeling, {Simulation} and
  {Optimization} for {Science} and {Technology}}}, no.~\bibinfo{number}{34} in
  \bibinfo{series}{Computational {Methods} in {Applied} {Sciences}},
  \bibinfo{pages}{41--58} (\bibinfo{publisher}{Springer Netherlands},
  \bibinfo{year}{2014}).
\newblock
  \urlprefix\url{http://link.springer.com/chapter/10.1007/978-94-017-9054-3_3}.
\newblock \bibinfo{note}{DOI: 10.1007/978-94-017-9054-3\_3}.

\bibitem{chen_new_2015}
\bibinfo{author}{Chen, B.}, \bibinfo{author}{Zeng, W.}, \bibinfo{author}{Lin,
  Y.} \& \bibinfo{author}{Zhang, D.}
\newblock \bibinfo{title}{A {New} {Local} {Search}-{Based} {Multiobjective}
  {Optimization} {Algorithm}}.
\newblock \emph{\bibinfo{journal}{IEEE Transactions on Evolutionary
  Computation}} \textbf{\bibinfo{volume}{19}}, \bibinfo{pages}{50--73}
  (\bibinfo{year}{2015}).

\bibitem{cai_fast_2000}
\bibinfo{author}{Cai, W.} \& \bibinfo{author}{Yu, T.}
\newblock \bibinfo{title}{Fast {Calculations} of {Dyadic} {Green}'s {Functions}
  for {Electromagnetic} {Scattering} in a {Multilayered} {Medium}}.
\newblock \emph{\bibinfo{journal}{Journal of Computational Physics}}
  \textbf{\bibinfo{volume}{165}}, \bibinfo{pages}{1--21}
  (\bibinfo{year}{2000}).
\newblock \urlprefix\url{http://dx.doi.org/10.1006/jcph.2000.6583}.

\bibitem{paulus_accurate_2000}
\bibinfo{author}{Paulus, M.}, \bibinfo{author}{Gay-Balmaz, P.} \&
  \bibinfo{author}{Martin, O. J.~F.}
\newblock \bibinfo{title}{Accurate and efficient computation of the {Green}'s
  tensor for stratified media}.
\newblock \emph{\bibinfo{journal}{Phys. Rev. E}} \textbf{\bibinfo{volume}{62}},
  \bibinfo{pages}{5797--5807} (\bibinfo{year}{2000}).
\newblock \urlprefix\url{http://link.aps.org/doi/10.1103/PhysRevE.62.5797}.

\bibitem{girard_shaping_2008}
\bibinfo{author}{Girard, C.}, \bibinfo{author}{Dujardin, E.},
  \bibinfo{author}{Baffou, G.} \& \bibinfo{author}{Quidant, R.}
\newblock \bibinfo{title}{Shaping and manipulation of light fields with
  bottom-up plasmonic structures}.
\newblock \emph{\bibinfo{journal}{New J. Phys.}} \textbf{\bibinfo{volume}{10}},
  \bibinfo{pages}{105016} (\bibinfo{year}{2008}).
\newblock \urlprefix\url{http://iopscience.iop.org/1367-2630/10/10/105016}.

\bibitem{malitson_interspecimen_1965}
\bibinfo{author}{Malitson, I.~H.}
\newblock \bibinfo{title}{Interspecimen {Comparison} of the {Refractive}
  {Index} of {Fused} {Silica}}.
\newblock \emph{\bibinfo{journal}{Journal of the Optical Society of America}}
  \textbf{\bibinfo{volume}{55}}, \bibinfo{pages}{1205} (\bibinfo{year}{1965}).
\newblock
  \urlprefix\url{https://www.osapublishing.org/abstract.cfm?URI=josa-55-10-1205}.

\bibitem{chaumet_simulation_2009}
\bibinfo{author}{Chaumet, P.~C.} \& \bibinfo{author}{Sentenac, A.}
\newblock \bibinfo{title}{Simulation of light scattering by multilayer
  cross-gratings with the coupled dipole method}.
\newblock \emph{\bibinfo{journal}{Journal of Quantitative Spectroscopy and
  Radiative Transfer}} \textbf{\bibinfo{volume}{110}},
  \bibinfo{pages}{409--414} (\bibinfo{year}{2009}).
\newblock
  \urlprefix\url{http://www.sciencedirect.com/science/article/pii/S0022407308002707}.

\bibitem{gallinet_electromagnetic_2009}
\bibinfo{author}{Gallinet, B.} \& \bibinfo{author}{Martin, O. J.~F.}
\newblock \bibinfo{title}{Electromagnetic {Scattering} of {Finite} and
  {Infinite} 3d {Lattices} in {Polarizable} {Backgrounds}}.
\newblock \emph{\bibinfo{journal}{Theoretical And Computational Nanophotonics
  (Tacona-Photonics 2009)}} \textbf{\bibinfo{volume}{1176}},
  \bibinfo{pages}{63--65} (\bibinfo{year}{2009}).
\newblock \urlprefix\url{http://infoscience.epfl.ch/record/164792}.

\end{thebibliography}

\end{document}